\documentclass[%
reprint,
superscriptaddress,
 amsmath,amssymb,
 aps,
]{revtex4-1}

\usepackage{graphicx}
\usepackage{dsfont}
\usepackage{dcolumn}
\usepackage{bm}
\usepackage{lipsum}
\usepackage{color}
\usepackage[normalem]{ulem}
\usepackage{simplewick}
\usepackage{qcircuit}
\usepackage{braket}
\usepackage{float}
\usepackage{multirow}
\usepackage{tikz}
\usepackage[utf8]{inputenc}

\usepackage{braket}
\usepackage{subfig}

\begin{document}

\title{Non-adiabatic molecular quantum dynamics with quantum computers}

\author{Pauline J. Ollitrault}
\affiliation{IBM Quantum, IBM Research -- Zurich, S\"aumerstrasse 4, 8803 R\"uschlikon, Switzerland}
\affiliation{Laboratory of Physical Chemistry,
ETH Zurich, Vladimir-Prelog-Weg 2, 8093 Zurich, Switzerland}

\author{Guglielmo Mazzola}
\affiliation{IBM Quantum, IBM Research -- Zurich, S\"aumerstrasse 4, 8803 R\"uschlikon, Switzerland}

\author{Ivano Tavernelli}
\email{ita@zurich.ibm.com}
\affiliation{IBM Quantum, IBM Research -- Zurich, S\"aumerstrasse 4, 8803 R\"uschlikon, Switzerland}

\date{\today}

\begin{abstract}
    The theoretical investigation of 
    non-adiabatic 
    processes is hampered by the complexity of the coupled electron-nuclear dynamics beyond the Born-Oppenheimer approximation. 
    Classically, the simulation of such reactions is limited by the unfavourable scaling of the computational resources 
    as a function of the system size.
    While quantum computing exhibits proven quantum advantage for the simulation of real-time dynamics, the study of quantum algorithms for the description of non-adiabatic phenomena is still unexplored. 
    In this work, we propose a quantum algorithm for the simulation of fast non-adiabatic chemical processes 
    together with an initialization scheme for quantum hardware calculations.
    In particular, we introduce a first-quantization method for the time evolution of a wavepacket on two coupled harmonic potential energy surfaces (Marcus model). 
    In our approach, the computational resources scale polynomially in the system dimensions, opening up new avenues for the study of photophysical processes that are classically intractable. 
\end{abstract}

\maketitle

Fast non-adiabatic processes are ubiquitous in science as they are the foundation of photo-induced reactions spanning the fields of biology
~\cite{domcke1997theory,worth2004, domcke2012, schuurman2018, yarkony2001},
chemical engineering and material science~\cite{hagfeldt1995light,Tang1987}.
From an atomistic standpoint, non-adiabatic dynamics account for various interesting phenomena. These include internal conversion and inter-system crossings among Born-Oppenheimer (BO) potential energy surfaces (PESs), the Jahn-Teller effect~\cite{bersuker2013}, and vibrational assisted energy versus electron transfer~\cite{tretiak2020,tavernelli2015,RevModPhys.90.035003}. 
In molecular systems, non-adiabatic processes occur through the dynamical coupling between the electronic and vibrational nuclear degrees of freedom. They are characterized by the break-down of the BO approximation~\cite{worth2004beyond}. 
However, the simultaneous description of the dynamics of the electronic and nuclear wavefunctions poses severe limitations to the size of the systems that can be simulated and to the accuracy of the solutions.
From a theoretical standpoint, relentless efforts have been made to refine numerical methods to simulate non-adiabatic phenomena beyond the analytically solvable Landau-Zener model~\cite{zener1932}.
First numerical attempts evolved around a semi-classical solution of the problem, e.g., within the Wenzel-Kramers-Brillouin approximation~\cite{miller1970}, the Ehrenfest dynamics~\cite{ehrenfest1927} and trajectory surface hopping~\cite{tully1990}.
These approaches have been extended to the study of non-adiabatic effects in molecules and solid state systems using first principle electronic structure approaches for the BO PESs~\cite{tavernelli2015, martinez2018}. However, the use of classical and quantum trajectories hampers a correct description of quantum phenomena, such as wavepacket branching at avoided crossings, tunneling, and quantum coherence and decoherence effects.

More naturally, the quantum dynamics of the nuclear wavefunction can be represented as a wavepacket, especially in those regimes where dynamics cannot be faithfully described by classical, semi-classical or quantum trajectories~\cite{sun1997semiclassical}. 
To this end, the direct solution of the time-dependent Schr\"odinger equation for the nuclear degrees of freedom is required. 
However, due to the 
exponential scaling of the Hilbert space,
grid methods can only be applied to low dimensional model Hamiltonians while the use of basis functions is usually limited to a few nuclear degrees of freedom~\cite{kosloff1994propagation,garraway1995wave,huang2018time}.
State-of-the-art approaches, like the Multi Configuration Time-Dependent Hartree (MCTDH) method~\cite{meyer1990,beck2000multiconfiguration}, can routinely tackle up to ten dimensions~\cite{egorova2001modeling,meyer2003quantum}, but not without the use of approximations.
In fact, as MCTDH relies on a compact time-dependent basis set description, the integration becomes less accurate as the propagation time increases, or when the dynamics becomes chaotic. 
Further approximations have been introduced with the aim at improving the efficiency of the method. These include non-orthogonal Gaussian-based G-MCTDH~\cite{burghardt2008}, local coherent state approximation~\cite{martinazzo2006}, and multiple spawning~\cite{martinez1997}.

Quantum computers can in principle simulate real-time quantum dynamics with polynomial complexity in memory and execution time.
Indeed, the simulation of quantum physics with quantum computers has been proposed theoretically decades ago by Feynman~\cite{feynman1999}, and realized experimentally in the last years for electronic structure calculations~\cite{kandala2017, colless2018, Kandala2019, Nam2019, ollitrault2019}.
Of particular interest is the possibility to perform wavepacket dynamics simulations in real-space representation~\cite{zalka1998, wiesner1996} with a quantum computer.
Within this framework, the space is discretized in a mesh of $\mathcal{N}$ points, separated by a distance $\Delta x$, in each dimension, that requires order $\log_2(\mathcal{N})$ memory space in the quantum register. The accuracy of the dynamics is not bounded by any basis set limitation but rather by $\Delta x$.
A general procedure for the real time propagation of a quantum state on a quantum computer was first introduced by Kassal \textit{et al.}~\cite{kassal2008}. However, this work did not specify how to encode the potential and kinetic energy terms of the time evolution operator into a quantum circuit (i.e. into operations on the qubits). 
Benenti and Strini~\cite{benenti2008} and later Somma~\cite{somma2015} presented a more detailed implementation in the case of a single, one-dimensional, harmonic potential.
Additionally, a quantum circuit to implement a spin-boson model has also been devised in Ref.~\cite{macridin2018electron}. 
In this case, the dynamical bosonic degrees of freedom in the model are represented as a wavepacket evolving under the action of a harmonic oscillator Hamiltonian, where the displacement operators are coupled with the $\sigma^z$ operator(s) of the spin(s).
Finally, a different approach consists of encoding the bosonic modes directly in the hardware, exploiting the microwave resonators available in the device~\cite{ballester2012quantum,leppakangas2018quantum}.

In this Letter, we devise a quantum algorithm for the simulation of non-adiabatic processes 
using a real-space representation of the wavepacket.
While described in a one-dimensional case, the method can easily be extended to the study of the quantum dynamics of larger systems by implementing additional spatial dimensions in the qubit register. 
The scheme is applied to the investigation of the dynamics of the one-dimensional Marcus model defined by two coupled harmonic potential energy curves, for which we observe the characteristic modulation of the charge transfer rates when going from the \emph{normal} to the so-called \emph{inverted} Marcus regime~\cite{marcus1993electron}.
Finally, we discuss and demonstrate the initialization of the Gaussian wavepacket on a quantum register. \\


\begin{figure}
    \centering
    \includegraphics[width = 0.9 \columnwidth]{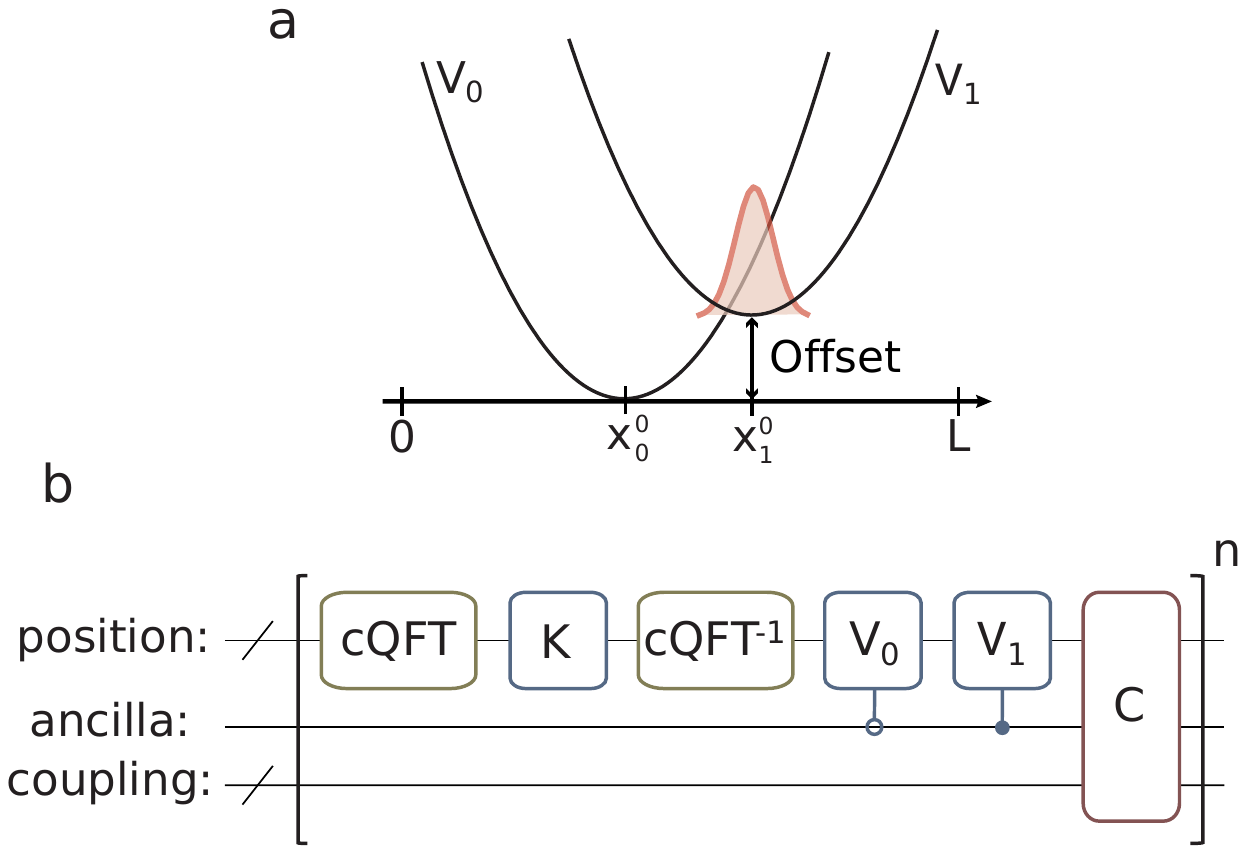}
    \caption{\textbf{a.} Graphical representation of the Marcus model. 
    \textbf{b.} Circuit for the time evolution of the wavepacket. The $K$, $V_i$ and $C$ blocks represent the time evolution operators for the kinetic, $i$th potential and coupling terms respectively.}
    \label{fig:model}
\end{figure}
\emph{The model.} The method aims at studying the dynamics of a wavepacket in several diabatic surfaces coupled through non-linear matrix elements in the first quantization formalism. 
For the sake of simplicity here we restrict the model to two one-dimensional diabatic curves. 
Note however that the generalization to multiple dimensions is straightforward.

The Hamiltonian of the system can then be written as 
\begin{equation}
    H = K \otimes \mathds{1}+  V_0 \otimes \ket{0}\bra{0} +  V_1 \otimes \ket{1}\bra{1} + C \otimes \sigma_x 
    \label{eq:hamiltonian}
\end{equation}
where $K = \frac{1}{2m}p^2$ is the kinetic energy operator of a particle with mass $m$ and momentum $p$, while $V_0$ and $V_1$ are the potentials of the first and the second diabatic curves respectively and are defined by functions of the position $x$. 
Likewise, the coupling operator $C$ is described by an arbitrary function of the position, $f(x)$.
An ancilla qubit, $\text{q}_N$, is entangled with the spacial register and controls the non-adiabatic dynamics accross the diabatic curves.
It is initialized in state $\ket{0}$ ($\ket{1}$) if the wavepacket at time $t=0$ is placed on the first (second) diabatic potential V$_0$ (V$_1$). 
For concreteness we specialize to the Marcus model~\cite{marcus1964chemical,marcus1993electron} which provides a simplified description of the electron-transfer reaction rate driven by collective outer and inner sphere coordinates~\cite{siders1981quantum}.
In this model, the potentials of Eq.~\eqref{eq:hamiltonian} are defined by $V_i = \frac{\omega_i ^2}{2m}(x-x^0_i)^2 + \Delta_rG^0_i$. The two harmonic potentials of the diabatic curves differ by an energy shift $\Delta_rG^0_i$, frequency $\omega_i$, equilibrium position $x^0_i$, and model the reactant and product states respectively. 
We call \textit{offset} the difference $\Delta_rG^0_1 - \Delta_rG^0_0$.
We use this setup in what follows and show its representation in Fig.~\ref{fig:model}a.\\

\emph{Resources scaling.} The position is encoded in the qubit register as $x = j\times \Delta x$ where $j$ is an integer which binary representation is encoded in the basis states of $N = \log_2(\mathcal{N})$ qubits.
Thus, in general,  we  require $d~\log_2(\mathcal{N})$ qubits to store the total wavefunction of the wavepacket in $d$ dimensions. 
An ancillary register of size $\lceil \log_2(\kappa) \rceil$ is needed to describe dynamics involving up to $\kappa$ diabatic potential energy surfaces.
In the specific model considered here, we only need one ancillary qubit for the propagation of the wavepacket in $V_0$ and $V_1$ coupled through the non-adiabatic coupling operator $C$.
Additionally, an extra qubit register is required to implement $C$ which size depends linearly on $N$ as well as on the shape of the coupling as explained later and in the Supplemental Material.\\

\emph{The time-evolution algorithm.}
The very first step of the dynamics resides in the initialization of the wavepacket in the quantum register.
In the interest of clarity, this step will be discussed in further detail at the end of this Letter and in the Supplemental Material. 
Then the wavepacket is propagated under the action of the real-time evolution operator such that $\ket{\Psi(t)} = e^{-\frac{i}{\hbar} H t}\ket{\Psi(t=0)}$,
using the Lie-Trotter-Suzuki product formulas~\cite{berry2007}. 
A quantum circuit for implementing the time evolution under the kinetic operator and harmonic potentials was presented in Refs.~\cite{somma2015, benenti2008} and is detailed in the Supplemental Material.
The same logic can be extended to potentials described by a polynomial function of the position. 
Note that to account for negative values of the momentum, a shift of $p_c = \Delta p  \mathcal{N}/2$, where $\Delta p = \frac{2 \pi}{\mathcal{N} \Delta x}$, is applied placing the zero momentum value at the center of the Brillouin zone. 
This choice implies the use of a centered Quantum Fourier Transform (cQFT) operator (see Supplemental Material for details) to implement the switch from the position to the momentum space.
While the quantum circuit for the kinetic part of the evolution can directly be applied on the first $N$ qubits, the potential parts must be controlled by the state of the ancilla qubit, $\text{q}_N$, such that the wavepacket evolves under the action of $e^{-iV_0t/n}$ and $e^{-iV_1t/n}$ when $q_N$ is in the $\ket{0}$ and $\ket{1}$ state respectively (here $n$ is the number of Trotter steps). 

One of the main methodological novelties of this work resides in the encoding of the coupling operator which acts as
\begin{equation}
    e^{-iC\otimes\sigma_xt/n}\ket{x}\ket{q_N} = e^{-if(x)\sigma_x t/n}\ket{x}\ket{q_N}
\end{equation}
and thus corresponds to a rotation of the ancilla qubit around the $x$ axis by an angle $f(x)t/n$. 
The general approach consists in pre-computing a discretized function, $f(x)$, into additional qubits using quantum arithmetic. 
While the number of gates and the number of additional qubits scale exponentially with the inverse desired accuracy for a general function~\cite{mitarai2019},
the resources can be kept reasonably low by approximating $f(x)$ as a piecewise linear function~\cite{haner2018}. 
In the approach adopted here, the additional gates and ancilla qubits scale linearly with $N$ and with the number of pieces in the description of $f(x)$ (see Supplemental Material and Refs.~\cite{woerner2019, stamatopoulos2019}).
Crucially, we observe that accurate results can be obtained by including only few pieces in $f(x)$.
A graphical representation of the quantum circuit used to encode the dynamics is shown in Fig.~\ref{fig:model}b.\\


%
\begin{figure}
    \includegraphics[width = 1 \columnwidth]{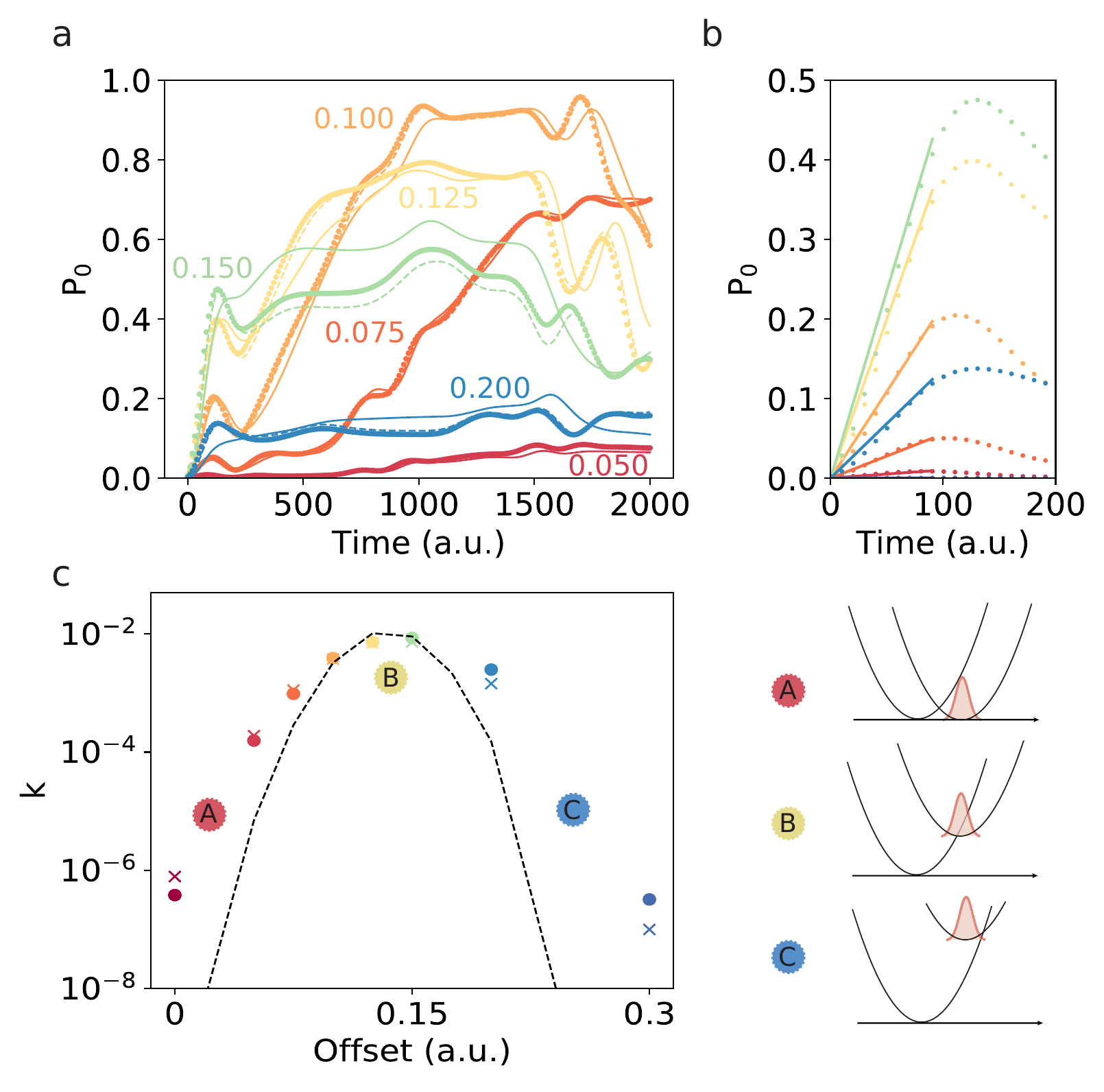}
    \caption{
    \textbf{a.} Time evolution of $P_0$ (see main text) obtained with our algorithm in classical simulations (dots), with the exact evolution with the reference coupling (full lines) and with the exact evolution with the approximate coupling (dashed lines). Curves at different offset values are displayed. 
    \textbf{b.} Linear fitting of the ten first steps of the evolution to approximate the rate constant.
    \textbf{c.} The approximated rate constant, $k$, as a function of the offset obtained with our algorithm (dots) and with the exact evolution with the reference coupling (crosses).
    The Marcus rates as calculated in the Supplemental Material are shown in dashed line for a qualitative comparison.
    The coloured stickers label the different charge transfer regions (A: normal regime, B: at reorganization energy and C: inverted region).}
    \label{fig:results}
\end{figure}
\begin{figure*}
    \includegraphics[width = 0.8 \textwidth]{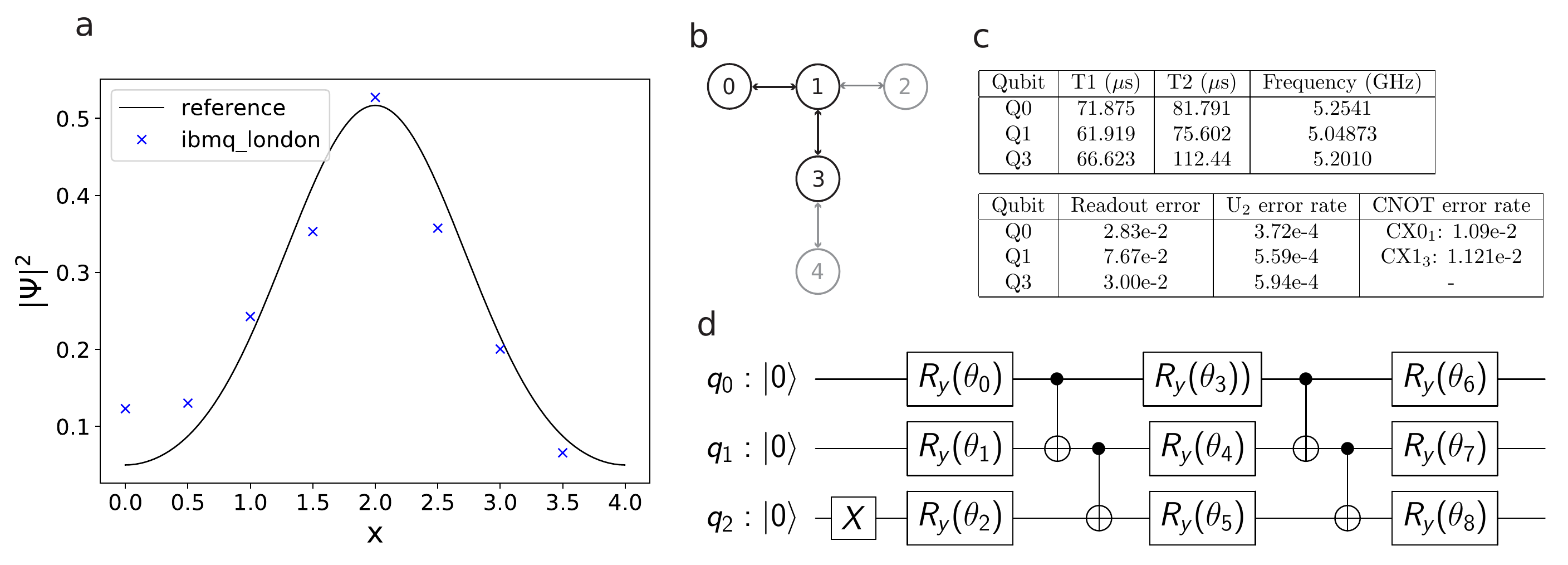}
    \caption{\textbf{a.} Initialization of a wavepacket in the \textit{ibmq\_london} 5-qubit device. The quantum circuit representing the discretized version of the reference wavepacket (black curve) is obtained from a classical simulation of the VQE algorithm. 
    \textbf{b.} Layout of the \textit{ibmq\_london} quantum computer. The qubits in black are the ones used to prepare the wavepacket. 
    \textbf{c.} Specifications of the qubit used to prepare the wavepacket.
    \textbf{d.} Quantum circuit for the preparartion of the wavepacket. The values of the optimized angles are $\theta_0 = -0.3383, \theta_1 = -1.5502,  \theta_2 = 2.3662,  \theta_3 = -0.6743, \theta_4 = -0.5438, \theta_5 = -2.0766,
 \theta_6 = -1.3717, \theta_7 = 0.3663, \theta_8 = -0.8286$}
    \label{fig:wp_init}
\end{figure*}
\emph{Rates in the Marcus model.}
The model parameters are chosen to represent typical molecular dynamics~\cite{yonehara2010} and the simulation conditions were optimized to guarantee the convergence of the results (see the Supplemental Material). 
Therefore, we choose to discretize a space of length $L=20$ using 8 qubits and we select a time-step of 10 a.u..
The non-adiabatic coupling term, which in the reference model is a Gaussian function, is approximated with a step function giving the best trade-off between accuracy and number of additional qubits that, in this case, amounts to 9 qubits.
The time evolution is performed for a total time of $T=2000$ a.u. and repeated for different values of the offset between the two harmonic potential energy curves. 

For the time being, we assume a Gaussian state preparation of the form (see below)~\cite{yonehara2010}
\begin{equation}
    \phi_0(x) = \big( \frac{1}{2\pi\delta^2}\big)^{1/4} e^{-\big(\frac{x-x_0}{2\delta}\big)^2} e^{ip_0(x-x_0)},
\end{equation}
at the center of the potential curve at the right, $V_1$, at $x_0 = 11.5$ (see Fig.~\ref{fig:model}a) with $\delta=1/3$. 
The initial momentum is set to $p_0=1$. This choice is motivated from the possibility to compare with the standard Marcus rate theory (see Supplemental Material). Note however that the qualitative behavior of the dynamics is not affected by the particular value of $p_0$. 

At each time step, the population fraction, $P_0$, in the product well $V_0$, is simply related to the expectation value of the ancilla qubit as  $P_0 = (\langle Z_{q_N} \rangle +1)/2$ (where $Z$ is the Pauli operator $\sigma_z$).
We run the dynamics for various offset values and show the time evolution of corresponding $P_0$ in Fig.~\ref{fig:results}a (dots), using a classical emulation of the corresponding quantum circuit. 
The exact evolution obtained with the original, Gaussian, coupling (full lines) and with its piecewise approximation (dashed lines) is also reported demonstrating the correct implementation of the algorithm (the small discrepancies are due to Trotter errors). 
Moreover, these results highlight that the piecewise linear approximation of the coupling function allows to recover the correct qualitative quantum dynamics. 
Note that the accuracy of the results is controllable as it can be systematically increased by improving the representation of the coupling function, as well as by reducing the Trotter step.
We calculate the initial rate constant, $k$, for each offset by taking the slope of the linear fit applied to the first ten steps of the dynamics as shown in Fig.~\ref{fig:results}b. 
The rates obtained from the quantum dynamics (dots) and from the exact dynamics with the exact, Gaussian, coupling (crosses) are in good agreement and are summarized in Fig.~\ref{fig:results}c.
As expected, the population transfer rate increases with the offset in the normal Marcus regime, reaching a maximum value when the offset is equal to the reorganization energy before decreasing again in the inverted region,
thus recovering the expected volcano shape predicted by Marcus theory (dashed line in Fig.~\ref{fig:results}c).
Note that discrepancies between the Marcus rates, calculated for a carefully chosen effective temperature (see the Supplemental Material) and our rate estimates are expected, as we are performing a closed-system dynamics.\\

\emph{Initial state preparation.}
To complete our presentation, we discuss an equally efficient initial state preparation method.
To this end, we rely on a more efficient Variational Quantum Eigensolver (VQE)~\cite{peruzzo2014,yung2014,mcclean2016,wang2019} approach instead of quantum arithmetic based methods~\cite{grover2002creating,kitaev2008wavefunction}.
As a proof of concept we show how to prepare a wavepacket defined as the ground state of the (arbitrarily chosen) Hamiltonian defined on a 3-qubit register (Supplemental Material for the detail).
The parametrized circuit comprises three layers of $R_y$ rotations intersected of 2 layers of CNOT gates as shown in Fig.~\ref{fig:wp_init}d. 
In the Supplemental Material we study the convergence of the VQE in presence of noise with the state-of-the-art optimizer and show that it requires an important number of optimization steps as well as error mitigation. 
Here the circuit can be optimized in a fully classical simulation of the VQE algorithm.
We employ this circuit on three qubits of the \textit{ibmq\_london} 5-qubit chip (see the hardware layout and the qubits specifications in Figs.~\ref{fig:wp_init}b and \ref{fig:wp_init}c respectively).
Since in this example $p_0=0$ (\textit{i.e.} the wavefunction is real) we simply display the modulo squared of the resulting wavefunction (obtained by measuring 8000 times in the position basis) in Fig.~\ref{fig:wp_init}a. We also show the reference Gaussian function demonstrating the initialization of the desired wavepacket in the quantum computer.\\

\emph{Conclusion.} 
We introduced a quantum algorithm to simulate the propagation of a nuclear wavepacket across $\kappa$ diabatic surfaces, featuring non-linear couplings.
The degrees of freedom are expressed in the first quantization formalism, as the position and momentum spaces are discretized and encoded in a \textit{position} quantum register.
Ancilla registers are used to encode the real time evolution of the population transfer between the $\kappa$ surfaces, and to realize the non-linear coupling operators.
The encoding of the problem is efficient in term of qubit resources, which scale logarithmically with the precision.
This impressive memory compression in storing the time-evolved wavefunction, represented in a systematically converging basis-set of a real space $\mathcal{N}$-point grid, readily realizes an exponential quantum advantage compared to classical algorithms.
As discussed, the proposed circuit to perform the coupled-time evolution only requires a polynomially scaling depth.
We demonstrate this approach to simulate the non-adiabatic dynamics of a wavepacket evolving in a Marcus model, consisting in two one-dimensional harmonic potentials shifted in energy by a variable offset.
This minimal model requires a feasible number of qubits (eighteen), so that the circuit can be classically emulated. The simulated dynamics are in excellent agreement with the exact propagation and we are able to observe the expected slowing down of the population transfer 
in the so-called \emph{inverted region}.
However, the circuit depth required to observe these dynamics greatly exceeds those currently feasible due to
the limited coherence time of present quantum hardware. 
Therefore, we limit the hardware demonstration to the first part of the algorithm \textit{i.e.} the Gaussian wavepacket initialization, on an IBM Q device.\\
As far as concerning the quantum resources (number of qubits), our algorithm can straightforwardly be extended to represent  polynomial potential energy surfaces in $d$-dimensions with a scaling $\mathcal{O}(d \log_2(\mathcal{N}))$. 
Hence, a quantum computer with $\sim165$ qubits would allow for the study of molecular systems characterized by up to 10 vibrational modes. 
Reaching the limits of classical simulations~\cite{capano2014,Tretiak_rev2015}, this approach will pave the way towards a better understanding of femto-chemistry processes, such as internal conversion and inter-system crossings, exciton formation and charge separation.\\

\emph{Acknowledgements.} The authors thank Julien Gacon, Almudena Carrera Vazquez and Stefan Woerner for useful discussions and acknowledge financial support from the Swiss National Science Foundation (SNF) through the grant No. 200021-179312.\\
IBM, the IBM logo, and ibm.com are trademarks of International Business Machines Corp., registered in many jurisdictions worldwide. Other product and service names might be trademarks of IBM or other companies. The current list of IBM trademarks is available at https://www.ibm.com/legal/copytrade.\\

%

\clearpage
\widetext
\begin{center}
\textbf{\large Supplementary Information for: Non-adiabatic molecular quantum dynamics with quantum computers}
\end{center}
%
\setcounter{equation}{0}
\setcounter{section}{0}
\setcounter{figure}{0}
\setcounter{table}{0}
\setcounter{page}{1}
\renewcommand{\theequation}{S\arabic{equation}}
\renewcommand{\thefigure}{S\arabic{figure}}
\renewcommand{\bibnumfmt}[1]{[#1]}
\renewcommand{\citenumfont}[1]{#1}
%
%
\section{Lie-Trotter decomposition}

The wavepacket is propagated under the action of the real-time evolution operator as
\begin{equation}
    \ket{\Psi(t)} = e^{-\frac{i}{\hbar} H t}\ket{\Psi(t=0)}.
\end{equation}
In practice the evolution operator is decomposed using Lie-Trotter-Suzuki product formulas~\cite{berry2007} to reduce the error arising from the fact that the kinetic part, $K$, of the Hamiltonian (acting in the momentum space) does not commute with the potential, $V$, and coupling, $C$, terms (in the position space). In this work we use the basic Lie-Trotter formula. Hence the evolution operator becomes
\begin{equation}
    \ket{\Psi(t)} = \bigg( e^{-\frac{i}{\hbar} K \frac{t}{n}} e^{-\frac{i}{\hbar} V \frac{t}{n}} e^{-\frac{i}{\hbar} C \frac{t}{n}} \bigg)^n \ket{\Psi(t=0)}.
\end{equation}

\section{Centered Quantum Fourier Transform}

The minimum of the momentum space is placed at the center of the Brillouin zone by applying a shift, $p_c = \frac{\Delta p \times \mathcal{N}}{2}$, to allow the momentum to take negative values. This must be taken into account in the QFT. In the case where the momentum space in centered exactly around the middle of the array we can simply add a X gate on the last qubit right before and after the QFT and QFT$^{-1}$ operations such that they undergo a cyclic permutation:
\begin{equation}
     \text{\textbf{cQFT}} =
    \begin{pmatrix}
     & & &1 & \cdots & 0\\
     &\cdots& & & \ddots & \\ 
     & & & 0 & \cdots & 1 \\
     1 & \cdots & 0 & & &\\
     & \ddots & & &\cdots&\\ 
     0 & \cdots & 1 &&& \\
    \end{pmatrix}
    \text{\textbf{QFT}}.
\end{equation}

\section{Quantum circuit for quadratic curve}
\begin{figure}
    \includegraphics[width = 0.8\textwidth]{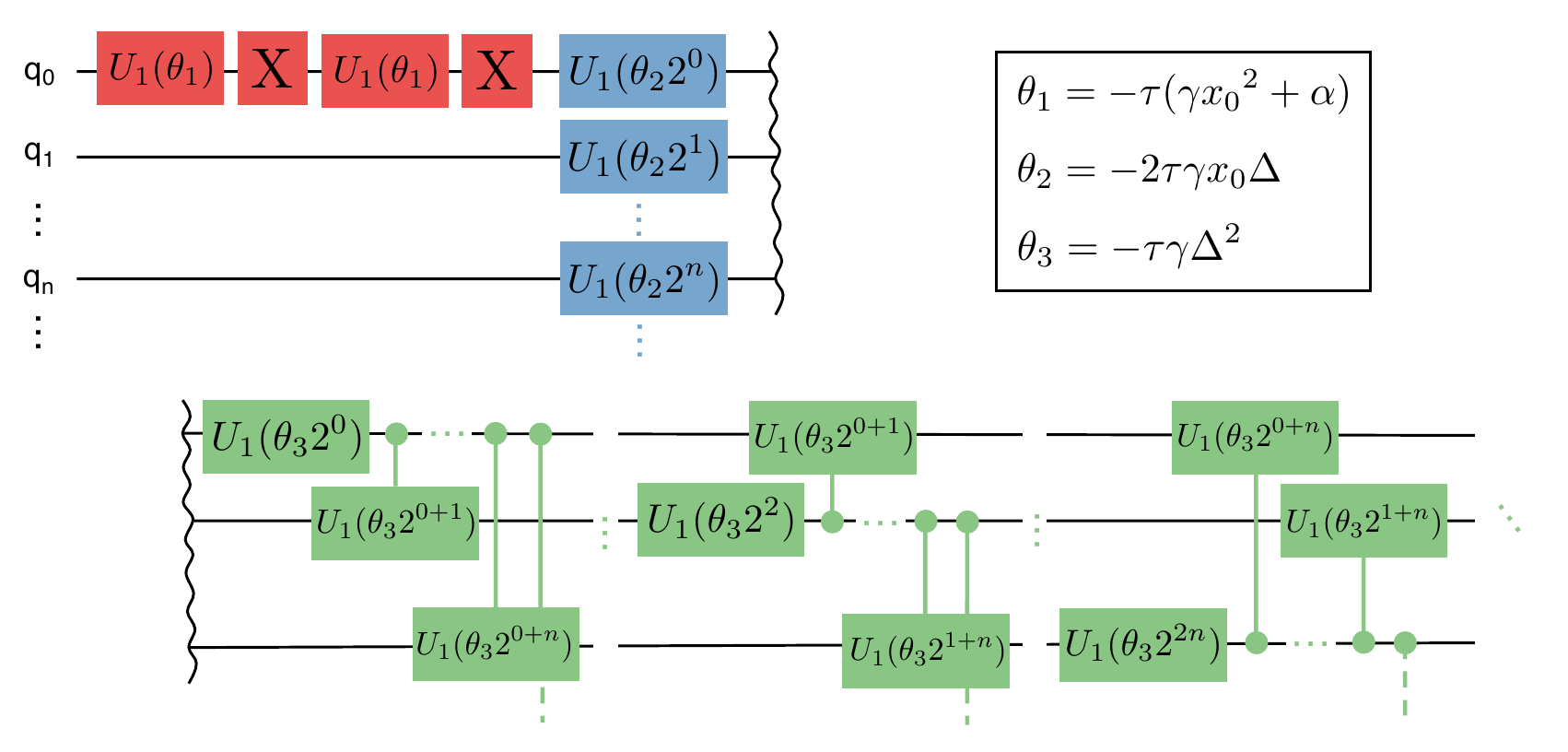}
    \caption{Quantum circuit implementing an operator of quadratic shape as defined by Eq.~\eqref{M_op}. The red, blue and green parts correspond to the $0^{\text{th}}$, $1^{\text{st}}$ and $2^{\text{nd}}$ order in the polynomial.}
    \label{fig:harmonic_circuit}
\end{figure}

The circuit for implementing any operator, $M$, of quadratic shape, i.e. which matrix elements are defined by
\begin{equation}
    M[m,n] = \left\{
    \begin{array}{ll}
        e^{-i\tau [\gamma(\Delta n + x_0)^2 + \alpha]} & \mbox{if } n=m, \\
        0 & \mbox{else}
    \end{array}
    \right.
    \label{M_op}
\end{equation}
is discussed in \cite{benenti2008} and depicted in Fig.~\ref{fig:harmonic_circuit}. The number of gates scales as $\mathcal{O}(N^2)$ with $N$ the number of qubits.

\section{Evaluation of a piecewise linear function}

\begin{figure}[h!]
    \centering
    \includegraphics[width = 0.7\textwidth]{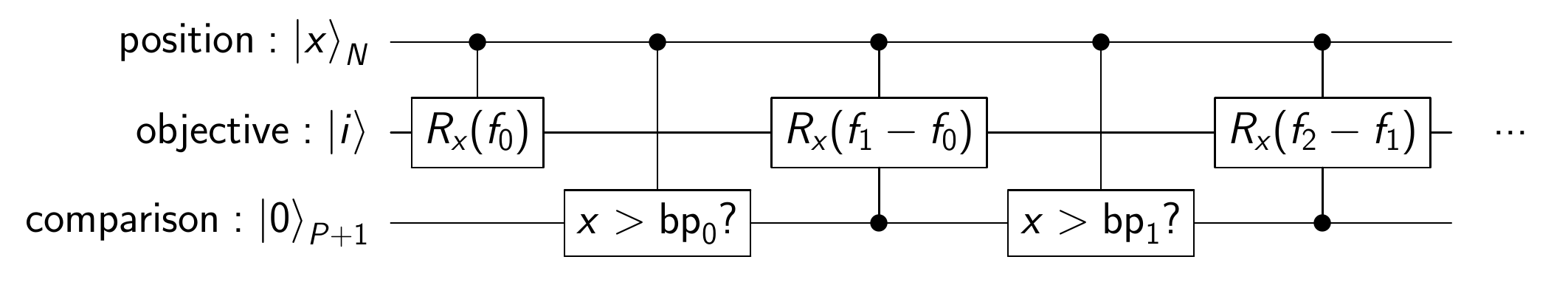}
    \caption{Quantum circuit implementing a rotation around the $x$ axis of the objective qubit which rotation angle is defined by a piecewise linear function of the position encoded in state of the position qubits. In this circuit `bp$_i$' stands for the $i^{th}$ breakpoint characterizing the piecewise linear function. }
    \label{fig:pwl_function_circuit}
\end{figure}

The piecewise linear function is defined as:
\begin{equation}
    f_{\text{pl}}(x)=\sum_{p=0}^P \alpha_p x + \beta_p
\end{equation}
where $P$ is total the number of pieces in the function.
A rotation of the type $e^{-i\tau (\alpha_p x + \beta_p) \sigma_x}$ can be applied on an \textit{objective} qubit, $q_{\text{obj}}$, in a similar way as presented in the previous section, by discretizing the variable $x$ as $x = \sum_{i=0}^{N-1}2^jk_j$ in the states $k_j$ of $N$ qubits (called \textit{position} qubits). 
The $P$ pieces of the function are delimited by $P-1$ breakpoints. The circuit for implementing the full operator $e^{-i\tau f_{\text{pl}}(x) \sigma_x}$ also requires additional \textit{comparison} qubits which role is clarified below. 
The number of \textit{comparison} qubits is $N_c = N + P - 2$ while the number of gates scales as $\mathcal{O}(PN)$~\cite{stamatopoulos2019}.
The algorithm works as follows. 
\begin{enumerate}
    \item Apply R$_x(\beta_0)$ to $q_{\text{obj}}$. 
    \item Apply CR$_x(\alpha_0 2^n)$ to $q_{\text{obj}}$ controlled by $q_n$ for each qubit $q_n$ in the position register.
    \item Check if $x>0^{th}$ break point using the comparison qubits and encode this information in the state of one comparison qubit $q_>$.
    \item Apply CR$_x(\beta_1 - \beta_0)$ to $q_{\text{obj}}$ controlled by $q_>$.
    \item Apply CR$_x((\alpha_1-\alpha_0) 2^n)$ to $q_{\text{obj}}$ controlled by $q_n$ for each qubit $q_n$ in the position register as well as by $q_>$.
    \item Repeat step 3, 4 and 5 for each piece $p$ of $f_{\text{pl}}(x)$ replacing the $0$ and $1$ indexes by $p-1$ and $p$ respectively. 
\end{enumerate}
For clarity we also display the corresponding circuit in Fig.~\ref{fig:pwl_function_circuit}.

\section{Parameters setting}

\begin{figure*}
    \centering
    \includegraphics[width = 0.7\textwidth]{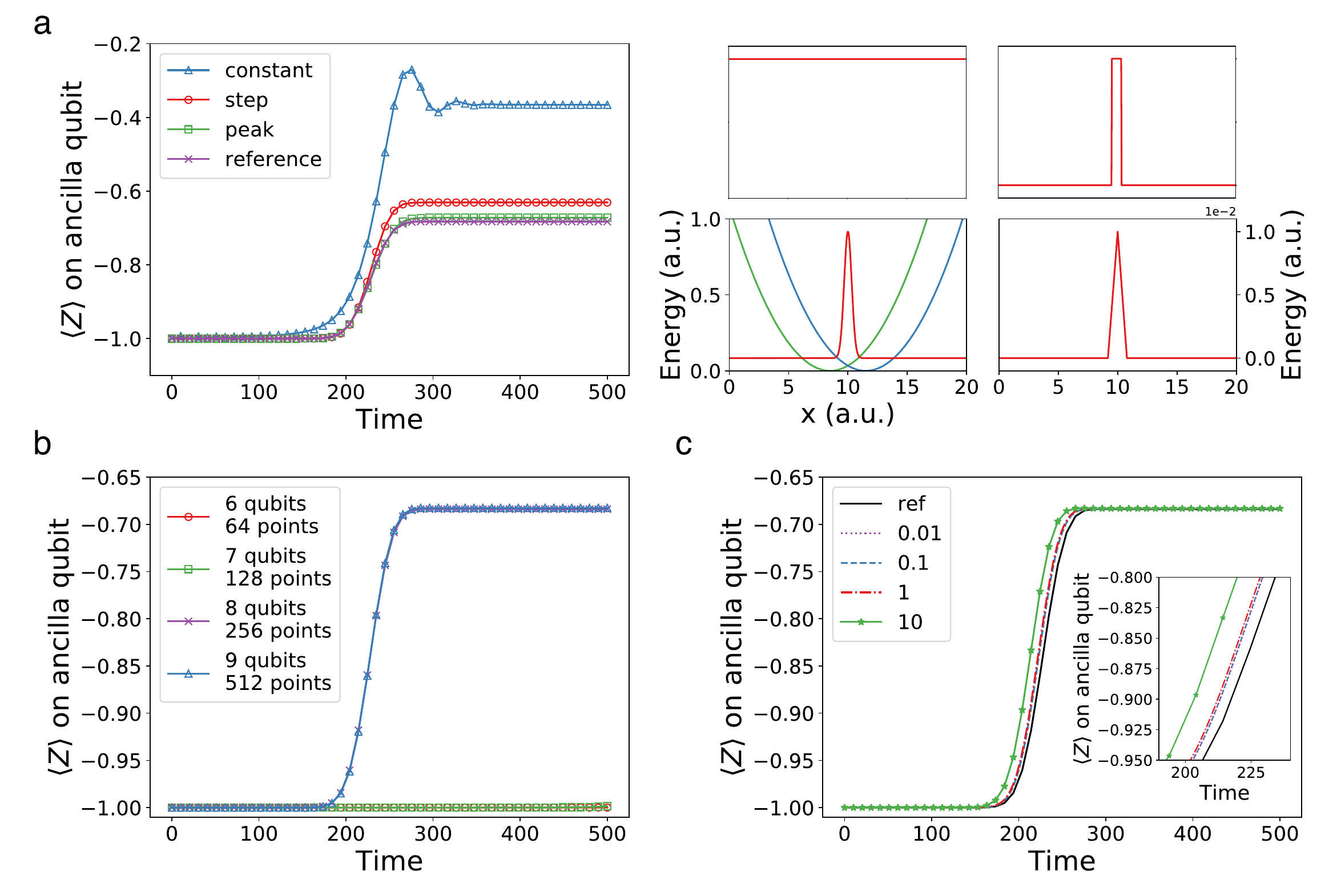}
    \caption{\textbf{a.} Left: Expectation value of the state of the ancilla qubit measured in the Z basis as a function of time for different coupling shapes. The evolution is obtained by applying the exact time evolution operator. Right: Shape of the four different coupling functions.
    \textbf{b.} Same as a. with reference coupling but for different number of qubits \textit{i.e.} discretization points.
    \textbf{c.} Expectation value of the state of the ancilla qubit measured in the Z basis as a function of time for different time steps. In this case the evolution is obtained by applying the Trotterized time evolution operator. The inset is a zoom on a given region of the time evolution.}
    \label{fig:preliminary_study}
\end{figure*}

This section is devoted to a preliminary study of the coupling shape, the space discretization and the Trotter error in order to extract reasonable parameters for the quantum evolution of the wavepacket. 
The non-adiabatic model~\cite{yonehara2010} employed in this work comprises two harmonic potential functions described by $V_1=\gamma(R-\alpha)^2$ and $V_2=\gamma(R+\alpha)^2$ with $\alpha=1.5$, $\beta=5.0$ and $\gamma=0.015$. The coupling potential is given by $V_c=\mu \exp{(-\beta x^2)}$ with $\mu=0.01$. The reduced mass is 1818.18 a.u. The length, $L$, of the box is fixed to 20. The wavefunction is initialized to a full quantum wavepacket,
\begin{equation}
    \phi_0(x) = \big( \frac{1}{2\pi\delta^2}\big)^{1/4} \exp\big(-\big(\frac{x-x_0}{2\delta}\big)^2\big) \exp(ip_0(x-x_0)),
\end{equation}
with $x_0=L/2+4$, $p_0=-30$ and $\delta=1/3$. 

First, we approximate the coupling shape with a piecewise linear function of the position.  
We try different coupling shapes shown on the right side of Fig.~\ref{fig:preliminary_study}a: a constant coupling (one piece), a step (three pieces) and a peak (four pieces).
The breaking points of the step and peak shapes are chosen to preserve the area under the reference curve given by 
\begin{equation}
    \int e^{-\alpha x^2} = \sqrt{\frac{\pi}{\alpha}}.
\end{equation}
The reference coupling shape is shown together with the two potential energy curves on the lower left panel. 
The initial wavepacket is propagated with the exact time evolution (no Trotter error) and the various coupling shapes. The number of qubit for the discretization of the position and momentum space is fixed to 9. The expectation value of the ancilla qubit measured in the Z basis is displayed in Fig.~\ref{fig:preliminary_study}a (left) and show both the step and peek coupling shapes provide qualitatively correct results. 
Although the peak shaped coupling leads to more accurate results, we choose to work with the step shape as it allows to save one qubit while remaining in good accuracy. 

As a second step, we evolve (using state-vector simulations) the wavepacket with the reference coupling shape and vary the number of qubits in the quantum register (and therefore the size of the grid's spacing). 
For the system under study, we find that a minimum of 8 qubits, equivalent to 256 grid points, is needed to reproduce faithfully the desired dynamics (see Fig.~\ref{fig:preliminary_study}b).

Finally we monitor the Trotter error. To do so, the number of qubits is kept fixed to 8, while the time step is varied. In Fig.~\ref{fig:preliminary_study}c, we show that qualitatively correct results can be obtained with a time step as large as 10 a.u., value that we chose for the production runs.

\section{Initialization of the wavepacket}

The initial wavepacket is prepared in the quantum computer by applying the corresponding quantum circuit. This quantum circuit can be found with the VQE algorithm~\cite{peruzzo2014,yung2014,mcclean2016,wang2019}. 
At each iteration of the VQE, the energy, $E$, of the trial wavefunction is calculated in the following way,
\begin{align}
    & E_{\text{pot}} = \frac{1}{N_{\text{shots}}}\sum_{j=0}^{\mathcal{N}}\frac{m\omega^2}{2} N_{\text{counts}}(j) (j\times\Delta x - x_0)^2 \\
    & E_{\text{kin}} = \frac{1}{N_{\text{shots}}}\sum_{j=0}^{\mathcal{N}}\frac{1}{2m} N_{\text{counts}}(j) (j\times\Delta p - p_0)^2 
\end{align}
and
\begin{equation}
    E = E_{\text{pot}} + E_{\text{kin}}
\end{equation}
where $E_{\text{pot}}$ and $E_{\text{kin}}$ are the potential and kinetic energy respectively. $N_{\text{shots}}$ is the total number of measurements done on the quantum computer to obtain the statistics, per basis. $N_{\text{counts}}(j)$ (with $0 \le N_{\text{counts}}(j) \le N_{\text{shots}} $, $\sum_j N_{\text{counts}}(j) = N_{\text{shots}}$) is the number of measurement that collapsed onto the qubit basis state corresponding to the binary representation of integer $j$. 
For the potential energy term the counts are obtained by measuring in the position basis where measurements can straightly be applied whereas the kinetic term requires applying a QFT beforehand to ensure that measurements are done in the momentum basis.

\begin{figure}
    \centering
    \includegraphics[width = 0.9\textwidth]{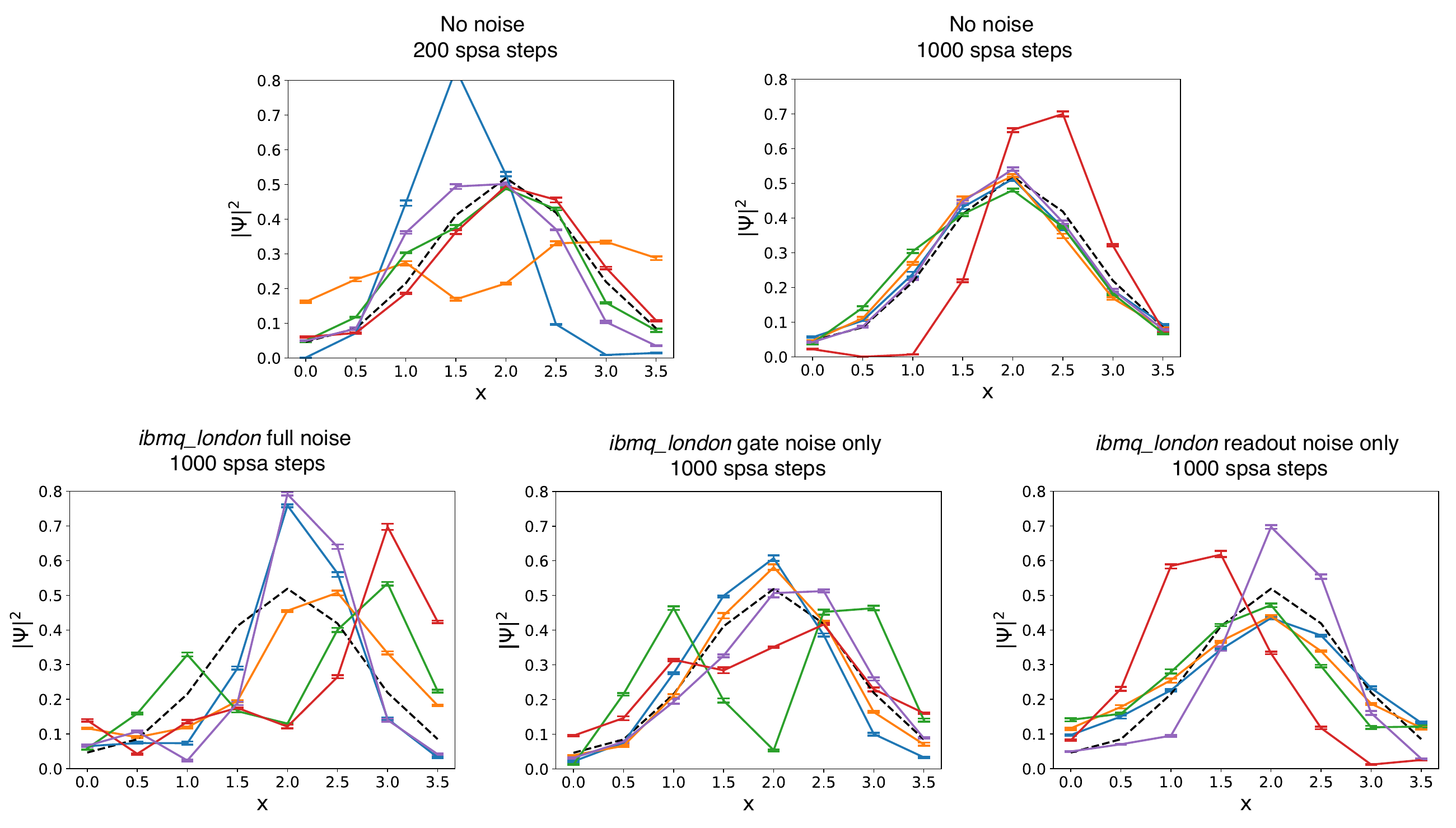}
    \caption{Results of the five VQE experiments realized on a classical simulator. The error bars are obtained by repeating 10 times the measurement of the VQE resulting circuit. The black dashed line represent the expected $|\Psi|^2$.}
    \label{fig:vqe_study_of_wp_init}
\end{figure}
We tested the ability of preparing the desired wavepacket with VQE for a 3-qubit system. 
In this example the wavepacket is described as the ground state of the Hamiltonian $H = 1/2\,p^2 + 1/2\,(x - 2)^2$. The size of the box is $L=4$. 
The trial wavefunction \textit{Ansatz} is shown in Fig.~\ref{fig:wp_init}d of the main text. 
At the beginning of the VQE, the quantum state is initialized as a (discrete) Dirac delta function $\delta(x-2)$, namely the bit string $|100\rangle$.
We run the VQE algorithm classically in a quantum simulator where the quantum measurements are simulated by projecting the statevector on the basis states 8000 times. 
Moreover, a noise model can be added to the classical simulator in order to simulate the gate noise (depolarization and thermal relaxation) and readout error characteristic of the available quantum computers. 
In what follows we employ a noise model based on the properties of the \textit{ibm\_q\_london} 5-qubit device. The layout and the properties of the exploited qubits are given in Fig.~\ref{fig:wp_init}b and c of the main text. 
We use the state-of-the-art optimizer for noisy optimization problems namely the Simultaneous Perturbation Stochastic Approximation (SPSA)\cite{spall2000}. 
We run five different experiments. Because the optimization process is stochastic, each experiment is composed of five VQE simulations.
The results are displayed in Fig.~\ref{fig:vqe_study_of_wp_init}.
In the first two experiments we completely turn off the noise model to study the bare convergence of the optimization. In the first case, the VQEs are stopped after 200 SPSA steps while in the second they are ended at 1000 steps. 
In the first case none of the VQEs converge whereas in the second case four out of five did showing that even for a circuit as shallow as the one employed for this task (3 qubits, 9 variational parameters and 4 CNOT gates) the convergence of the SPSA algorithm requires a consequent number of iterations. 
We then turn on the noise model fully. In this case, none of the VQEs are able to recover the desired Gaussian shape after 1000 SPSA steps leading to the conclusion that error mitigation is imperative. 
Finally we realize two experiments by keeping only 1. the gate noise and 2. the readout errors. In both cases three out of five VQEs were able to reasonably approach the desired shape showing that by reducing the noise (e.g. with error mitigation) the SPSA optimizer may be robust enough to approximate the ground state within reasonable accuracy. 

\section{Marcus rates}

The Marcus rate constants used in the main text are given by~\cite{marcus1956,nitzan2006}
\begin{equation}
    k_M = \frac{2\pi |V|^2}{\hbar}\sqrt{\frac{\beta}{4\pi\lambda}}\exp{\big(-\frac{\beta(\lambda-\epsilon)^2}{4\lambda}\big)}
\end{equation}
where $V$ is the electronic coupling, $\lambda$ the reorganization energy, $\epsilon$ the offset and $\hbar$ the Planck's constant and $\beta = 1/k_BT$ the reciprocal of Boltzmann’s constant, $k_B$, times the temperature $T$. 

Seeking for a qualitative comparison between the rates resulting from the quantum dynamics and the Marcus rates we need to determine a sensible $\beta$ value to insert in this equation.
Since our quantum dynamics does not include dissipation or any coupling with the environment, the temperature is not present as an explicit parameter in our setting.
The route we adopt here is to extract an effective temperature by a careful initialization of the wavepacket.

\begin{figure}
    \centering
    \includegraphics[width=0.7\textwidth]{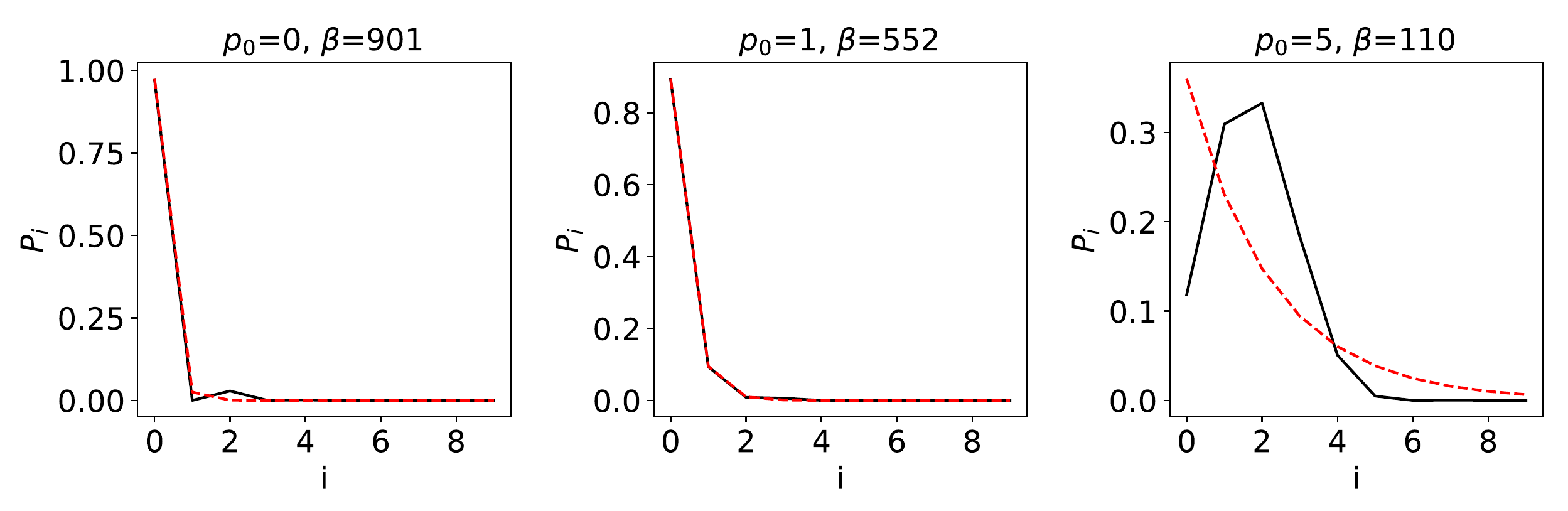}
    \caption{Density of states given by the projection of the Gaussian wavepacket on the eigenstates of $H_{HO_1}$ (black line) and by the best fit of the Boltzmann distribution (dashed red line) given by the indicated $\beta$ for $p_0=0$, $p_0=1$ and $p_0=5$.}
    \label{fig:boltzmann_fit}
\end{figure}

To do so, we look for an initial position and momentum, $x_0$ and $p_0$, for which the density of states given by the initial Gaussian wavepacket, $\Psi(x_0, p_0)$, can be described by a Boltzmann distribution such that
\begin{equation}
    P_i = |\braket{i|\Psi(x_0,p_0)}|^2 = \frac{\exp{(-\beta E_i)}}{\sum_{j=0}^{\mathcal{N}-1}\exp{(-\beta E_j)}}
\end{equation}
where $\ket{i}$ are the eigenstates of the harmonic oscillator defined by $H_{HO_1} = \frac{p^2}{2m} + V_1$ and $E_i$ is the energy of state $\ket{i}$. 
We choose $x_0=11.5$, the center of $V_1$ and calculate the $P_i$ for several values of $p_0$. We then fit the resulting distribution to a Boltzmann distribution to extract a value for $\beta$ (see Fig.~\ref{fig:boltzmann_fit}). 
We find that for a value of $p_0=1$ the density of states can be well represented by a Boltzmann distribution defined by $\beta=552$. 
This set our choice for the $p_0$ parameter in our simulations and for the $\beta$ parameter in the Marcus rate theory to compare with.
We notice obviously that the quantum dynamics can be performed with any choice of $p_0$, but this specific choice is made to enable a qualitative comparison with Marcus theory.
Therefore, we run the quantum dynamics initializing the Gaussian wavepacket with $x_0=11.5$ and $p_0=1$ and compute the Marcus rates with $\beta=552$.


\begin{thebibliography}{59}%
\makeatletter
\providecommand \@ifxundefined [1]{%
 \@ifx{#1\undefined}
}%
\providecommand \@ifnum [1]{%
 \ifnum #1\expandafter \@firstoftwo
 \else \expandafter \@secondoftwo
 \fi
}%
\providecommand \@ifx [1]{%
 \ifx #1\expandafter \@firstoftwo
 \else \expandafter \@secondoftwo
 \fi
}%
\providecommand \natexlab [1]{#1}%
\providecommand \enquote  [1]{``#1''}%
\providecommand \bibnamefont  [1]{#1}%
\providecommand \bibfnamefont [1]{#1}%
\providecommand \citenamefont [1]{#1}%
\providecommand \href@noop [0]{\@secondoftwo}%
\providecommand \href [0]{\begingroup \@sanitize@url \@href}%
\providecommand \@href[1]{\@@startlink{#1}\@@href}%
\providecommand \@@href[1]{\endgroup#1\@@endlink}%
\providecommand \@sanitize@url [0]{\catcode `\\12\catcode `\$12\catcode
  `\&12\catcode `\#12\catcode `\^12\catcode `\_12\catcode `\%12\relax}%
\providecommand \@@startlink[1]{}%
\providecommand \@@endlink[0]{}%
\providecommand \url  [0]{\begingroup\@sanitize@url \@url }%
\providecommand \@url [1]{\endgroup\@href {#1}{\urlprefix }}%
\providecommand \urlprefix  [0]{URL }%
\providecommand \Eprint [0]{\href }%
\providecommand \doibase [0]{http://dx.doi.org/}%
\providecommand \selectlanguage [0]{\@gobble}%
\providecommand \bibinfo  [0]{\@secondoftwo}%
\providecommand \bibfield  [0]{\@secondoftwo}%
\providecommand \translation [1]{[#1]}%
\providecommand \BibitemOpen [0]{}%
\providecommand \bibitemStop [0]{}%
\providecommand \bibitemNoStop [0]{.\EOS\space}%
\providecommand \EOS [0]{\spacefactor3000\relax}%
\providecommand \BibitemShut  [1]{\csname bibitem#1\endcsname}%
\let\auto@bib@innerbib\@empty
\bibitem [{\citenamefont {Domcke}\ and\ \citenamefont
  {Stock}(1997)}]{domcke1997theory}%
  \BibitemOpen
  \bibfield  {author} {\bibinfo {author} {\bibfnamefont {W.}~\bibnamefont
  {Domcke}}\ and\ \bibinfo {author} {\bibfnamefont {G.}~\bibnamefont {Stock}},\
  }\href@noop {} {\bibfield  {journal} {\bibinfo  {journal} {Advances in
  Chemical Physics}\ }\textbf {\bibinfo {volume} {100}},\ \bibinfo {pages} {1}
  (\bibinfo {year} {1997})}\BibitemShut {NoStop}%
\bibitem [{\citenamefont {Worth}\ and\ \citenamefont
  {Cederbaum}(2004{\natexlab{a}})}]{worth2004}%
  \BibitemOpen
  \bibfield  {author} {\bibinfo {author} {\bibfnamefont {G.~A.}\ \bibnamefont
  {Worth}}\ and\ \bibinfo {author} {\bibfnamefont {L.~S.}\ \bibnamefont
  {Cederbaum}},\ }\href@noop {} {\bibfield  {journal} {\bibinfo  {journal}
  {Annual Review of Physical Chemistry}\ }\textbf {\bibinfo {volume} {55}},\
  \bibinfo {pages} {127} (\bibinfo {year} {2004}{\natexlab{a}})}\BibitemShut
  {NoStop}%
\bibitem [{\citenamefont {Domcke}\ and\ \citenamefont
  {Yarkony}(2012)}]{domcke2012}%
  \BibitemOpen
  \bibfield  {author} {\bibinfo {author} {\bibfnamefont {W.}~\bibnamefont
  {Domcke}}\ and\ \bibinfo {author} {\bibfnamefont {D.~R.}\ \bibnamefont
  {Yarkony}},\ }\href@noop {} {\bibfield  {journal} {\bibinfo  {journal}
  {Annual Review of Physical Chemistry}\ }\textbf {\bibinfo {volume} {63}},\
  \bibinfo {pages} {325} (\bibinfo {year} {2012})}\BibitemShut {NoStop}%
\bibitem [{\citenamefont {Schuurman}\ and\ \citenamefont
  {Stolow}(2018)}]{schuurman2018}%
  \BibitemOpen
  \bibfield  {author} {\bibinfo {author} {\bibfnamefont {M.~S.}\ \bibnamefont
  {Schuurman}}\ and\ \bibinfo {author} {\bibfnamefont {A.}~\bibnamefont
  {Stolow}},\ }\href@noop {} {\bibfield  {journal} {\bibinfo  {journal} {Annual
  Review of Physical Chemistry}\ }\textbf {\bibinfo {volume} {69}},\ \bibinfo
  {pages} {427} (\bibinfo {year} {2018})}\BibitemShut {NoStop}%
\bibitem [{\citenamefont {Yarkony}(2001)}]{yarkony2001}%
  \BibitemOpen
  \bibfield  {author} {\bibinfo {author} {\bibfnamefont {D.~R.}\ \bibnamefont
  {Yarkony}},\ }\href@noop {} {\bibfield  {journal} {\bibinfo  {journal} {The
  Journal of Physical Chemistry A}\ }\textbf {\bibinfo {volume} {105}},\
  \bibinfo {pages} {6277} (\bibinfo {year} {2001})}\BibitemShut {NoStop}%
\bibitem [{\citenamefont {Hagfeldt}\ and\ \citenamefont
  {Graetzel}(1995)}]{hagfeldt1995light}%
  \BibitemOpen
  \bibfield  {author} {\bibinfo {author} {\bibfnamefont {A.}~\bibnamefont
  {Hagfeldt}}\ and\ \bibinfo {author} {\bibfnamefont {M.}~\bibnamefont
  {Graetzel}},\ }\href@noop {} {\bibfield  {journal} {\bibinfo  {journal}
  {Chemical Reviews}\ }\textbf {\bibinfo {volume} {95}},\ \bibinfo {pages} {49}
  (\bibinfo {year} {1995})}\BibitemShut {NoStop}%
\bibitem [{\citenamefont {Tang}\ and\ \citenamefont
  {Vanslyke}(1987)}]{Tang1987}%
  \BibitemOpen
  \bibfield  {author} {\bibinfo {author} {\bibfnamefont {C.~W.}\ \bibnamefont
  {Tang}}\ and\ \bibinfo {author} {\bibfnamefont {S.~A.}\ \bibnamefont
  {Vanslyke}},\ }\href@noop {} {\bibfield  {journal} {\bibinfo  {journal}
  {Applied Physics Letters}\ }\textbf {\bibinfo {volume} {51}},\ \bibinfo
  {pages} {913} (\bibinfo {year} {1987})}\BibitemShut {NoStop}%
\bibitem [{\citenamefont {Bersuker}(2013)}]{bersuker2013}%
  \BibitemOpen
  \bibfield  {author} {\bibinfo {author} {\bibfnamefont {I.}~\bibnamefont
  {Bersuker}},\ }\href@noop {} {\emph {\bibinfo {title} {The Jahn-Teller effect
  and vibronic interactions in modern chemistry}}}\ (\bibinfo  {publisher}
  {Springer Science \& Business Media},\ \bibinfo {year} {2013})\BibitemShut
  {NoStop}%
\bibitem [{\citenamefont {Nelson}\ \emph {et~al.}(2020)\citenamefont {Nelson},
  \citenamefont {White}, \citenamefont {Bjorgaard}, \citenamefont {Sifain},
  \citenamefont {Zhang}, \citenamefont {Nebgen}, \citenamefont
  {Fernandez-Alberti}, \citenamefont {Mozyrsky}, \citenamefont {Roitberg},\
  and\ \citenamefont {Tretiak}}]{tretiak2020}%
  \BibitemOpen
  \bibfield  {author} {\bibinfo {author} {\bibfnamefont {T.}~\bibnamefont
  {Nelson}}, \bibinfo {author} {\bibfnamefont {A.}~\bibnamefont {White}},
  \bibinfo {author} {\bibfnamefont {J.}~\bibnamefont {Bjorgaard}}, \bibinfo
  {author} {\bibfnamefont {A.}~\bibnamefont {Sifain}}, \bibinfo {author}
  {\bibfnamefont {Y.}~\bibnamefont {Zhang}}, \bibinfo {author} {\bibfnamefont
  {B.}~\bibnamefont {Nebgen}}, \bibinfo {author} {\bibfnamefont
  {S.}~\bibnamefont {Fernandez-Alberti}}, \bibinfo {author} {\bibfnamefont
  {D.}~\bibnamefont {Mozyrsky}}, \bibinfo {author} {\bibfnamefont
  {A.}~\bibnamefont {Roitberg}}, \ and\ \bibinfo {author} {\bibfnamefont
  {S.}~\bibnamefont {Tretiak}},\ }\href {\doibase 10.1021/acs.chemrev.9b00447}
  {\bibfield  {journal} {\bibinfo  {journal} {Chemical Reviews}\ }\textbf
  {\bibinfo {volume} {120}} (\bibinfo {year} {2020}),\
  10.1021/acs.chemrev.9b00447}\BibitemShut {NoStop}%
\bibitem [{\citenamefont {Tavernelli}(2015)}]{tavernelli2015}%
  \BibitemOpen
  \bibfield  {author} {\bibinfo {author} {\bibfnamefont {I.}~\bibnamefont
  {Tavernelli}},\ }\href@noop {} {\bibfield  {journal} {\bibinfo  {journal}
  {Accounts of Chemical Research}\ }\textbf {\bibinfo {volume} {48}},\ \bibinfo
  {pages} {792} (\bibinfo {year} {2015})}\BibitemShut {NoStop}%
\bibitem [{\citenamefont {Jang}\ and\ \citenamefont
  {Mennucci}(2018)}]{RevModPhys.90.035003}%
  \BibitemOpen
  \bibfield  {author} {\bibinfo {author} {\bibfnamefont {S.~J.}\ \bibnamefont
  {Jang}}\ and\ \bibinfo {author} {\bibfnamefont {B.}~\bibnamefont
  {Mennucci}},\ }\href {\doibase 10.1103/RevModPhys.90.035003} {\bibfield
  {journal} {\bibinfo  {journal} {Rev. Mod. Phys.}\ }\textbf {\bibinfo {volume}
  {90}},\ \bibinfo {pages} {035003} (\bibinfo {year} {2018})}\BibitemShut
  {NoStop}%
\bibitem [{\citenamefont {Worth}\ and\ \citenamefont
  {Cederbaum}(2004{\natexlab{b}})}]{worth2004beyond}%
  \BibitemOpen
  \bibfield  {author} {\bibinfo {author} {\bibfnamefont {G.~A.}\ \bibnamefont
  {Worth}}\ and\ \bibinfo {author} {\bibfnamefont {L.~S.}\ \bibnamefont
  {Cederbaum}},\ }\href@noop {} {\bibfield  {journal} {\bibinfo  {journal}
  {Annual Review of Physical Chemistry}\ }\textbf {\bibinfo {volume} {55}},\
  \bibinfo {pages} {127} (\bibinfo {year} {2004}{\natexlab{b}})}\BibitemShut
  {NoStop}%
\bibitem [{\citenamefont {Zener}(1932)}]{zener1932}%
  \BibitemOpen
  \bibfield  {author} {\bibinfo {author} {\bibfnamefont {C.}~\bibnamefont
  {Zener}},\ }\href@noop {} {\bibfield  {journal} {\bibinfo  {journal} {Proc.
  R. Soc. A}\ }\textbf {\bibinfo {volume} {137}},\ \bibinfo {pages} {696}
  (\bibinfo {year} {1932})}\BibitemShut {NoStop}%
\bibitem [{\citenamefont {Miller}(1970)}]{miller1970}%
  \BibitemOpen
  \bibfield  {author} {\bibinfo {author} {\bibfnamefont {W.~H.}\ \bibnamefont
  {Miller}},\ }\href@noop {} {\bibfield  {journal} {\bibinfo  {journal} {J.
  Chem. Phys}\ }\textbf {\bibinfo {volume} {53}},\ \bibinfo {pages} {3578}
  (\bibinfo {year} {1970})}\BibitemShut {NoStop}%
\bibitem [{\citenamefont {Ehrenfest}(1927)}]{ehrenfest1927}%
  \BibitemOpen
  \bibfield  {author} {\bibinfo {author} {\bibfnamefont {P.}~\bibnamefont
  {Ehrenfest}},\ }\href@noop {} {\bibfield  {journal} {\bibinfo  {journal}
  {Zeitschrift f{\"u}r Physik}\ }\textbf {\bibinfo {volume} {45}},\ \bibinfo
  {pages} {455} (\bibinfo {year} {1927})}\BibitemShut {NoStop}%
\bibitem [{\citenamefont {Tully}(1990)}]{tully1990}%
  \BibitemOpen
  \bibfield  {author} {\bibinfo {author} {\bibfnamefont {J.~C.}\ \bibnamefont
  {Tully}},\ }\href@noop {} {\bibfield  {journal} {\bibinfo  {journal} {The
  Journal of Chemical Physics}\ }\textbf {\bibinfo {volume} {93}},\ \bibinfo
  {pages} {1061} (\bibinfo {year} {1990})}\BibitemShut {NoStop}%
\bibitem [{\citenamefont {Curchod}\ and\ \citenamefont
  {Martínez}(2018)}]{martinez2018}%
  \BibitemOpen
  \bibfield  {author} {\bibinfo {author} {\bibfnamefont {B.~F.~E.}\
  \bibnamefont {Curchod}}\ and\ \bibinfo {author} {\bibfnamefont {T.~J.}\
  \bibnamefont {Martínez}},\ }\href@noop {} {\bibfield  {journal} {\bibinfo
  {journal} {Chemical Reviews}\ }\textbf {\bibinfo {volume} {118}},\ \bibinfo
  {pages} {3305} (\bibinfo {year} {2018})}\BibitemShut {NoStop}%
\bibitem [{\citenamefont {Sun}\ and\ \citenamefont
  {Miller}(1997)}]{sun1997semiclassical}%
  \BibitemOpen
  \bibfield  {author} {\bibinfo {author} {\bibfnamefont {X.}~\bibnamefont
  {Sun}}\ and\ \bibinfo {author} {\bibfnamefont {W.~H.}\ \bibnamefont
  {Miller}},\ }\href@noop {} {\bibfield  {journal} {\bibinfo  {journal} {The
  Journal of Chemical Physics}\ }\textbf {\bibinfo {volume} {106}},\ \bibinfo
  {pages} {6346} (\bibinfo {year} {1997})}\BibitemShut {NoStop}%
\bibitem [{\citenamefont {Kosloff}(1994)}]{kosloff1994propagation}%
  \BibitemOpen
  \bibfield  {author} {\bibinfo {author} {\bibfnamefont {R.}~\bibnamefont
  {Kosloff}},\ }\href@noop {} {\bibfield  {journal} {\bibinfo  {journal}
  {Annual Review of Physical Chemistry}\ }\textbf {\bibinfo {volume} {45}},\
  \bibinfo {pages} {145} (\bibinfo {year} {1994})}\BibitemShut {NoStop}%
\bibitem [{\citenamefont {Garraway}\ and\ \citenamefont
  {Suominen}(1995)}]{garraway1995wave}%
  \BibitemOpen
  \bibfield  {author} {\bibinfo {author} {\bibfnamefont {B.~M.}\ \bibnamefont
  {Garraway}}\ and\ \bibinfo {author} {\bibfnamefont {K.-A.}\ \bibnamefont
  {Suominen}},\ }\href@noop {} {\bibfield  {journal} {\bibinfo  {journal}
  {Reports on Progress in Physics}\ }\textbf {\bibinfo {volume} {58}},\
  \bibinfo {pages} {365} (\bibinfo {year} {1995})}\BibitemShut {NoStop}%
\bibitem [{\citenamefont {Huang}\ \emph {et~al.}(2018)\citenamefont {Huang},
  \citenamefont {Liu}, \citenamefont {Zhang},\ and\ \citenamefont
  {Krems}}]{huang2018time}%
  \BibitemOpen
  \bibfield  {author} {\bibinfo {author} {\bibfnamefont {J.}~\bibnamefont
  {Huang}}, \bibinfo {author} {\bibfnamefont {S.}~\bibnamefont {Liu}}, \bibinfo
  {author} {\bibfnamefont {D.~H.}\ \bibnamefont {Zhang}}, \ and\ \bibinfo
  {author} {\bibfnamefont {R.~V.}\ \bibnamefont {Krems}},\ }\href@noop {}
  {\bibfield  {journal} {\bibinfo  {journal} {Physical Review Letters}\
  }\textbf {\bibinfo {volume} {120}},\ \bibinfo {pages} {143401} (\bibinfo
  {year} {2018})}\BibitemShut {NoStop}%
\bibitem [{\citenamefont {Meyer}\ \emph {et~al.}(1990)\citenamefont {Meyer},
  \citenamefont {Manthe},\ and\ \citenamefont {Cederbaum}}]{meyer1990}%
  \BibitemOpen
  \bibfield  {author} {\bibinfo {author} {\bibfnamefont {H.-D.}\ \bibnamefont
  {Meyer}}, \bibinfo {author} {\bibfnamefont {U.}~\bibnamefont {Manthe}}, \
  and\ \bibinfo {author} {\bibfnamefont {L.~S.}\ \bibnamefont {Cederbaum}},\
  }\href@noop {} {\bibfield  {journal} {\bibinfo  {journal} {Chemical Physics
  Letters}\ }\textbf {\bibinfo {volume} {165}},\ \bibinfo {pages} {73}
  (\bibinfo {year} {1990})}\BibitemShut {NoStop}%
\bibitem [{\citenamefont {Beck}\ \emph {et~al.}(2000)\citenamefont {Beck},
  \citenamefont {J{\"a}ckle}, \citenamefont {Worth},\ and\ \citenamefont
  {Meyer}}]{beck2000multiconfiguration}%
  \BibitemOpen
  \bibfield  {author} {\bibinfo {author} {\bibfnamefont {M.~H.}\ \bibnamefont
  {Beck}}, \bibinfo {author} {\bibfnamefont {A.}~\bibnamefont {J{\"a}ckle}},
  \bibinfo {author} {\bibfnamefont {G.~A.}\ \bibnamefont {Worth}}, \ and\
  \bibinfo {author} {\bibfnamefont {H.-D.}\ \bibnamefont {Meyer}},\ }\href@noop
  {} {\bibfield  {journal} {\bibinfo  {journal} {Physics Reports}\ }\textbf
  {\bibinfo {volume} {324}},\ \bibinfo {pages} {1} (\bibinfo {year}
  {2000})}\BibitemShut {NoStop}%
\bibitem [{\citenamefont {Egorova}\ \emph {et~al.}(2001)\citenamefont
  {Egorova}, \citenamefont {K{\"u}hl},\ and\ \citenamefont
  {Domcke}}]{egorova2001modeling}%
  \BibitemOpen
  \bibfield  {author} {\bibinfo {author} {\bibfnamefont {D.}~\bibnamefont
  {Egorova}}, \bibinfo {author} {\bibfnamefont {A.}~\bibnamefont {K{\"u}hl}}, \
  and\ \bibinfo {author} {\bibfnamefont {W.}~\bibnamefont {Domcke}},\
  }\href@noop {} {\bibfield  {journal} {\bibinfo  {journal} {Chemical Physics}\
  }\textbf {\bibinfo {volume} {268}},\ \bibinfo {pages} {105} (\bibinfo {year}
  {2001})}\BibitemShut {NoStop}%
\bibitem [{\citenamefont {Meyer}\ and\ \citenamefont
  {Worth}(2003)}]{meyer2003quantum}%
  \BibitemOpen
  \bibfield  {author} {\bibinfo {author} {\bibfnamefont {H.-D.}\ \bibnamefont
  {Meyer}}\ and\ \bibinfo {author} {\bibfnamefont {G.~A.}\ \bibnamefont
  {Worth}},\ }\href@noop {} {\bibfield  {journal} {\bibinfo  {journal}
  {Theoretical Chemistry Accounts}\ }\textbf {\bibinfo {volume} {109}},\
  \bibinfo {pages} {251} (\bibinfo {year} {2003})}\BibitemShut {NoStop}%
\bibitem [{\citenamefont {Burghardt}\ \emph {et~al.}(2008)\citenamefont
  {Burghardt}, \citenamefont {Giri},\ and\ \citenamefont
  {Worth}}]{burghardt2008}%
  \BibitemOpen
  \bibfield  {author} {\bibinfo {author} {\bibfnamefont {I.}~\bibnamefont
  {Burghardt}}, \bibinfo {author} {\bibfnamefont {K.}~\bibnamefont {Giri}}, \
  and\ \bibinfo {author} {\bibnamefont {Worth}},\ }\href@noop {} {\bibfield
  {journal} {\bibinfo  {journal} {J. Chem. Phys.}\ }\textbf {\bibinfo {volume}
  {129}},\ \bibinfo {pages} {174104} (\bibinfo {year} {2008})}\BibitemShut
  {NoStop}%
\bibitem [{\citenamefont {Martinazzo}\ \emph {et~al.}(2006)\citenamefont
  {Martinazzo}, \citenamefont {Nest}, \citenamefont {Saalfrank},\ and\
  \citenamefont {Tantardini}}]{martinazzo2006}%
  \BibitemOpen
  \bibfield  {author} {\bibinfo {author} {\bibfnamefont {R.}~\bibnamefont
  {Martinazzo}}, \bibinfo {author} {\bibfnamefont {M.}~\bibnamefont {Nest}},
  \bibinfo {author} {\bibfnamefont {P.}~\bibnamefont {Saalfrank}}, \ and\
  \bibinfo {author} {\bibfnamefont {G.~F.}\ \bibnamefont {Tantardini}},\
  }\href@noop {} {\bibfield  {journal} {\bibinfo  {journal} {J. Chem. Phys.}\
  }\textbf {\bibinfo {volume} {125}},\ \bibinfo {pages} {194102} (\bibinfo
  {year} {2006})}\BibitemShut {NoStop}%
\bibitem [{\citenamefont {Mart{\'\i}nez}\ and\ \citenamefont
  {Levine}(1997)}]{martinez1997}%
  \BibitemOpen
  \bibfield  {author} {\bibinfo {author} {\bibfnamefont {T.~J.}\ \bibnamefont
  {Mart{\'\i}nez}}\ and\ \bibinfo {author} {\bibfnamefont {R.~D.}\ \bibnamefont
  {Levine}},\ }\href@noop {} {\bibfield  {journal} {\bibinfo  {journal}
  {Journal of the Chemical Society, Faraday Transactions}\ }\textbf {\bibinfo
  {volume} {93}},\ \bibinfo {pages} {941} (\bibinfo {year} {1997})}\BibitemShut
  {NoStop}%
\bibitem [{\citenamefont {Feynman}(1999)}]{feynman1999}%
  \BibitemOpen
  \bibfield  {author} {\bibinfo {author} {\bibfnamefont {R.~P.}\ \bibnamefont
  {Feynman}},\ }\href@noop {} {\bibfield  {journal} {\bibinfo  {journal}
  {International Journal of Theoretical Physics}\ }\textbf {\bibinfo {volume}
  {21}} (\bibinfo {year} {1999})}\BibitemShut {NoStop}%
\bibitem [{\citenamefont {Kandala}\ \emph {et~al.}(2017)\citenamefont
  {Kandala}, \citenamefont {Mezzacapo}, \citenamefont {Temme}, \citenamefont
  {Takita}, \citenamefont {Brink}, \citenamefont {Chow},\ and\ \citenamefont
  {Gambetta}}]{kandala2017}%
  \BibitemOpen
  \bibfield  {author} {\bibinfo {author} {\bibfnamefont {A.}~\bibnamefont
  {Kandala}}, \bibinfo {author} {\bibfnamefont {A.}~\bibnamefont {Mezzacapo}},
  \bibinfo {author} {\bibfnamefont {K.}~\bibnamefont {Temme}}, \bibinfo
  {author} {\bibfnamefont {M.}~\bibnamefont {Takita}}, \bibinfo {author}
  {\bibfnamefont {M.}~\bibnamefont {Brink}}, \bibinfo {author} {\bibfnamefont
  {J.~M.}\ \bibnamefont {Chow}}, \ and\ \bibinfo {author} {\bibfnamefont
  {J.~M.}\ \bibnamefont {Gambetta}},\ }\href@noop {} {\bibfield  {journal}
  {\bibinfo  {journal} {Nature}\ }\textbf {\bibinfo {volume} {549}},\ \bibinfo
  {pages} {242} (\bibinfo {year} {2017})}\BibitemShut {NoStop}%
\bibitem [{\citenamefont {Colless}\ \emph {et~al.}(2018)\citenamefont
  {Colless}, \citenamefont {Ramasesh}, \citenamefont {Dahlen}, \citenamefont
  {Blok}, \citenamefont {Kimchi-Schwartz}, \citenamefont {McClean},
  \citenamefont {Carter}, \citenamefont {De~Jong},\ and\ \citenamefont
  {Siddiqi}}]{colless2018}%
  \BibitemOpen
  \bibfield  {author} {\bibinfo {author} {\bibfnamefont {J.~I.}\ \bibnamefont
  {Colless}}, \bibinfo {author} {\bibfnamefont {V.~V.}\ \bibnamefont
  {Ramasesh}}, \bibinfo {author} {\bibfnamefont {D.}~\bibnamefont {Dahlen}},
  \bibinfo {author} {\bibfnamefont {M.~S.}\ \bibnamefont {Blok}}, \bibinfo
  {author} {\bibfnamefont {M.}~\bibnamefont {Kimchi-Schwartz}}, \bibinfo
  {author} {\bibfnamefont {J.}~\bibnamefont {McClean}}, \bibinfo {author}
  {\bibfnamefont {J.}~\bibnamefont {Carter}}, \bibinfo {author} {\bibfnamefont
  {W.}~\bibnamefont {De~Jong}}, \ and\ \bibinfo {author} {\bibfnamefont
  {I.}~\bibnamefont {Siddiqi}},\ }\href@noop {} {\bibfield  {journal} {\bibinfo
   {journal} {Physical Review X}\ }\textbf {\bibinfo {volume} {8}},\ \bibinfo
  {pages} {011021} (\bibinfo {year} {2018})}\BibitemShut {NoStop}%
\bibitem [{\citenamefont {Kandala}\ \emph {et~al.}(2019)\citenamefont
  {Kandala}, \citenamefont {Temme}, \citenamefont {C{\'{o}}rcoles},
  \citenamefont {Mezzacapo}, \citenamefont {Chow},\ and\ \citenamefont
  {Gambetta}}]{Kandala2019}%
  \BibitemOpen
  \bibfield  {author} {\bibinfo {author} {\bibfnamefont {A.}~\bibnamefont
  {Kandala}}, \bibinfo {author} {\bibfnamefont {K.}~\bibnamefont {Temme}},
  \bibinfo {author} {\bibfnamefont {A.~D.}\ \bibnamefont {C{\'{o}}rcoles}},
  \bibinfo {author} {\bibfnamefont {A.}~\bibnamefont {Mezzacapo}}, \bibinfo
  {author} {\bibfnamefont {J.~M.}\ \bibnamefont {Chow}}, \ and\ \bibinfo
  {author} {\bibfnamefont {J.~M.}\ \bibnamefont {Gambetta}},\ }\href {\doibase
  10.1038/s41586-019-1040-7} {\bibfield  {journal} {\bibinfo  {journal}
  {Nature}\ }\textbf {\bibinfo {volume} {567}},\ \bibinfo {pages} {491}
  (\bibinfo {year} {2019})}\BibitemShut {NoStop}%
\bibitem [{\citenamefont {Nam}\ \emph {et~al.}(2019)\citenamefont {Nam},
  \citenamefont {Chen}, \citenamefont {Pisenti}, \citenamefont {Wright},
  \citenamefont {Delaney}, \citenamefont {Maslov}, \citenamefont {Brown},
  \citenamefont {Allen}, \citenamefont {Amini}, \citenamefont {Apisdorf} \emph
  {et~al.}}]{Nam2019}%
  \BibitemOpen
  \bibfield  {author} {\bibinfo {author} {\bibfnamefont {Y.}~\bibnamefont
  {Nam}}, \bibinfo {author} {\bibfnamefont {J.-S.}\ \bibnamefont {Chen}},
  \bibinfo {author} {\bibfnamefont {N.~C.}\ \bibnamefont {Pisenti}}, \bibinfo
  {author} {\bibfnamefont {K.}~\bibnamefont {Wright}}, \bibinfo {author}
  {\bibfnamefont {C.}~\bibnamefont {Delaney}}, \bibinfo {author} {\bibfnamefont
  {D.}~\bibnamefont {Maslov}}, \bibinfo {author} {\bibfnamefont {K.~R.}\
  \bibnamefont {Brown}}, \bibinfo {author} {\bibfnamefont {S.}~\bibnamefont
  {Allen}}, \bibinfo {author} {\bibfnamefont {J.~M.}\ \bibnamefont {Amini}},
  \bibinfo {author} {\bibfnamefont {J.}~\bibnamefont {Apisdorf}},  \emph
  {et~al.},\ }\href@noop {} {\bibfield  {journal} {\bibinfo  {journal} {arXiv
  preprint arXiv:1902.10171}\ } (\bibinfo {year} {2019})}\BibitemShut {NoStop}%
\bibitem [{\citenamefont {Ollitrault}\ \emph {et~al.}(2019)\citenamefont
  {Ollitrault}, \citenamefont {Kandala}, \citenamefont {Chen}, \citenamefont
  {Barkoutsos}, \citenamefont {Mezzacapo}, \citenamefont {Pistoia},
  \citenamefont {Sheldon}, \citenamefont {Woerner}, \citenamefont {Gambetta},\
  and\ \citenamefont {Tavernelli}}]{ollitrault2019}%
  \BibitemOpen
  \bibfield  {author} {\bibinfo {author} {\bibfnamefont {P.~J.}\ \bibnamefont
  {Ollitrault}}, \bibinfo {author} {\bibfnamefont {A.}~\bibnamefont {Kandala}},
  \bibinfo {author} {\bibfnamefont {C.-F.}\ \bibnamefont {Chen}}, \bibinfo
  {author} {\bibfnamefont {P.~K.}\ \bibnamefont {Barkoutsos}}, \bibinfo
  {author} {\bibfnamefont {A.}~\bibnamefont {Mezzacapo}}, \bibinfo {author}
  {\bibfnamefont {M.}~\bibnamefont {Pistoia}}, \bibinfo {author} {\bibfnamefont
  {S.}~\bibnamefont {Sheldon}}, \bibinfo {author} {\bibfnamefont
  {S.}~\bibnamefont {Woerner}}, \bibinfo {author} {\bibfnamefont
  {J.}~\bibnamefont {Gambetta}}, \ and\ \bibinfo {author} {\bibfnamefont
  {I.}~\bibnamefont {Tavernelli}},\ }\href@noop {} {\bibfield  {journal}
  {\bibinfo  {journal} {arXiv preprint arXiv:1910.12890}\ } (\bibinfo {year}
  {2019})}\BibitemShut {NoStop}%
\bibitem [{\citenamefont {Zalka}(1998)}]{zalka1998}%
  \BibitemOpen
  \bibfield  {author} {\bibinfo {author} {\bibfnamefont {C.}~\bibnamefont
  {Zalka}},\ }\href@noop {} {\bibfield  {journal} {\bibinfo  {journal}
  {Proceedings of the Royal Society of London. Series A: Mathematical, Physical
  and Engineering Sciences}\ }\textbf {\bibinfo {volume} {454}},\ \bibinfo
  {pages} {313} (\bibinfo {year} {1998})}\BibitemShut {NoStop}%
\bibitem [{\citenamefont {Wiesner}(1996)}]{wiesner1996}%
  \BibitemOpen
  \bibfield  {author} {\bibinfo {author} {\bibfnamefont {S.}~\bibnamefont
  {Wiesner}},\ }\href@noop {} {\bibfield  {journal} {\bibinfo  {journal} {arXiv
  preprint arXiv:quant-ph/9603028}\ } (\bibinfo {year} {1996})}\BibitemShut
  {NoStop}%
\bibitem [{\citenamefont {Kassal}\ \emph {et~al.}(2008)\citenamefont {Kassal},
  \citenamefont {Jordan}, \citenamefont {Love}, \citenamefont {Mohseni},\ and\
  \citenamefont {Aspuru-Guzik}}]{kassal2008}%
  \BibitemOpen
  \bibfield  {author} {\bibinfo {author} {\bibfnamefont {I.}~\bibnamefont
  {Kassal}}, \bibinfo {author} {\bibfnamefont {S.~P.}\ \bibnamefont {Jordan}},
  \bibinfo {author} {\bibfnamefont {P.~J.}\ \bibnamefont {Love}}, \bibinfo
  {author} {\bibfnamefont {M.}~\bibnamefont {Mohseni}}, \ and\ \bibinfo
  {author} {\bibfnamefont {A.}~\bibnamefont {Aspuru-Guzik}},\ }\href@noop {}
  {\bibfield  {journal} {\bibinfo  {journal} {Proceedings of the National
  Academy of Sciences}\ }\textbf {\bibinfo {volume} {105}},\ \bibinfo {pages}
  {18681} (\bibinfo {year} {2008})}\BibitemShut {NoStop}%
\bibitem [{\citenamefont {Benenti}\ and\ \citenamefont
  {Strini}(2008)}]{benenti2008}%
  \BibitemOpen
  \bibfield  {author} {\bibinfo {author} {\bibfnamefont {G.}~\bibnamefont
  {Benenti}}\ and\ \bibinfo {author} {\bibfnamefont {G.}~\bibnamefont
  {Strini}},\ }\href@noop {} {\bibfield  {journal} {\bibinfo  {journal}
  {American Journal of Physics}\ }\textbf {\bibinfo {volume} {76}},\ \bibinfo
  {pages} {657} (\bibinfo {year} {2008})}\BibitemShut {NoStop}%
\bibitem [{\citenamefont {Somma}(2015)}]{somma2015}%
  \BibitemOpen
  \bibfield  {author} {\bibinfo {author} {\bibfnamefont {R.~D.}\ \bibnamefont
  {Somma}},\ }\href@noop {} {\bibfield  {journal} {\bibinfo  {journal} {arXiv
  preprint arXiv:1503.06319v2}\ } (\bibinfo {year} {2015})}\BibitemShut
  {NoStop}%
\bibitem [{\citenamefont {Macridin}\ \emph {et~al.}(2018)\citenamefont
  {Macridin}, \citenamefont {Spentzouris}, \citenamefont {Amundson},\ and\
  \citenamefont {Harnik}}]{macridin2018electron}%
  \BibitemOpen
  \bibfield  {author} {\bibinfo {author} {\bibfnamefont {A.}~\bibnamefont
  {Macridin}}, \bibinfo {author} {\bibfnamefont {P.}~\bibnamefont
  {Spentzouris}}, \bibinfo {author} {\bibfnamefont {J.}~\bibnamefont
  {Amundson}}, \ and\ \bibinfo {author} {\bibfnamefont {R.}~\bibnamefont
  {Harnik}},\ }\href@noop {} {\bibfield  {journal} {\bibinfo  {journal}
  {Physical Review Letters}\ }\textbf {\bibinfo {volume} {121}},\ \bibinfo
  {pages} {110504} (\bibinfo {year} {2018})}\BibitemShut {NoStop}%
\bibitem [{\citenamefont {Ballester}\ \emph {et~al.}(2012)\citenamefont
  {Ballester}, \citenamefont {Romero}, \citenamefont {Garc{\'\i}a-Ripoll},
  \citenamefont {Deppe},\ and\ \citenamefont {Solano}}]{ballester2012quantum}%
  \BibitemOpen
  \bibfield  {author} {\bibinfo {author} {\bibfnamefont {D.}~\bibnamefont
  {Ballester}}, \bibinfo {author} {\bibfnamefont {G.}~\bibnamefont {Romero}},
  \bibinfo {author} {\bibfnamefont {J.~J.}\ \bibnamefont {Garc{\'\i}a-Ripoll}},
  \bibinfo {author} {\bibfnamefont {F.}~\bibnamefont {Deppe}}, \ and\ \bibinfo
  {author} {\bibfnamefont {E.}~\bibnamefont {Solano}},\ }\href@noop {}
  {\bibfield  {journal} {\bibinfo  {journal} {Physical Review X}\ }\textbf
  {\bibinfo {volume} {2}},\ \bibinfo {pages} {021007} (\bibinfo {year}
  {2012})}\BibitemShut {NoStop}%
\bibitem [{\citenamefont {Lepp{\"a}kangas}\ \emph {et~al.}(2018)\citenamefont
  {Lepp{\"a}kangas}, \citenamefont {Braum{\"u}ller}, \citenamefont {Hauck},
  \citenamefont {Reiner}, \citenamefont {Schwenk}, \citenamefont {Zanker},
  \citenamefont {Fritz}, \citenamefont {Ustinov}, \citenamefont {Weides},\ and\
  \citenamefont {Marthaler}}]{leppakangas2018quantum}%
  \BibitemOpen
  \bibfield  {author} {\bibinfo {author} {\bibfnamefont {J.}~\bibnamefont
  {Lepp{\"a}kangas}}, \bibinfo {author} {\bibfnamefont {J.}~\bibnamefont
  {Braum{\"u}ller}}, \bibinfo {author} {\bibfnamefont {M.}~\bibnamefont
  {Hauck}}, \bibinfo {author} {\bibfnamefont {J.-M.}\ \bibnamefont {Reiner}},
  \bibinfo {author} {\bibfnamefont {I.}~\bibnamefont {Schwenk}}, \bibinfo
  {author} {\bibfnamefont {S.}~\bibnamefont {Zanker}}, \bibinfo {author}
  {\bibfnamefont {L.}~\bibnamefont {Fritz}}, \bibinfo {author} {\bibfnamefont
  {A.~V.}\ \bibnamefont {Ustinov}}, \bibinfo {author} {\bibfnamefont
  {M.}~\bibnamefont {Weides}}, \ and\ \bibinfo {author} {\bibfnamefont
  {M.}~\bibnamefont {Marthaler}},\ }\href@noop {} {\bibfield  {journal}
  {\bibinfo  {journal} {Physical Review A}\ }\textbf {\bibinfo {volume} {97}},\
  \bibinfo {pages} {052321} (\bibinfo {year} {2018})}\BibitemShut {NoStop}%
\bibitem [{\citenamefont {Marcus}(1993)}]{marcus1993electron}%
  \BibitemOpen
  \bibfield  {author} {\bibinfo {author} {\bibfnamefont {R.~A.}\ \bibnamefont
  {Marcus}},\ }\href@noop {} {\bibfield  {journal} {\bibinfo  {journal}
  {Reviews of Modern Physics}\ }\textbf {\bibinfo {volume} {65}},\ \bibinfo
  {pages} {599} (\bibinfo {year} {1993})}\BibitemShut {NoStop}%
\bibitem [{\citenamefont {Marcus}(1964)}]{marcus1964chemical}%
  \BibitemOpen
  \bibfield  {author} {\bibinfo {author} {\bibfnamefont {R.~A.}\ \bibnamefont
  {Marcus}},\ }\href@noop {} {\bibfield  {journal} {\bibinfo  {journal} {Annual
  Review of Physical Chemistry}\ }\textbf {\bibinfo {volume} {15}},\ \bibinfo
  {pages} {155} (\bibinfo {year} {1964})}\BibitemShut {NoStop}%
\bibitem [{\citenamefont {Siders}\ and\ \citenamefont
  {Marcus}(1981)}]{siders1981quantum}%
  \BibitemOpen
  \bibfield  {author} {\bibinfo {author} {\bibfnamefont {P.}~\bibnamefont
  {Siders}}\ and\ \bibinfo {author} {\bibfnamefont {R.~A.}\ \bibnamefont
  {Marcus}},\ }\href@noop {} {\bibfield  {journal} {\bibinfo  {journal}
  {Journal of the American Chemical Society}\ }\textbf {\bibinfo {volume}
  {103}},\ \bibinfo {pages} {748} (\bibinfo {year} {1981})}\BibitemShut
  {NoStop}%
\bibitem [{\citenamefont {Berry}\ \emph {et~al.}(2007)\citenamefont {Berry},
  \citenamefont {Ahokas}, \citenamefont {Cleve},\ and\ \citenamefont
  {Sanders}}]{berry2007}%
  \BibitemOpen
  \bibfield  {author} {\bibinfo {author} {\bibfnamefont {D.~W.}\ \bibnamefont
  {Berry}}, \bibinfo {author} {\bibfnamefont {G.}~\bibnamefont {Ahokas}},
  \bibinfo {author} {\bibfnamefont {R.}~\bibnamefont {Cleve}}, \ and\ \bibinfo
  {author} {\bibfnamefont {B.~C.}\ \bibnamefont {Sanders}},\ }\href@noop {}
  {\bibfield  {journal} {\bibinfo  {journal} {Communications in Mathematical
  Physics}\ }\textbf {\bibinfo {volume} {270}},\ \bibinfo {pages} {359}
  (\bibinfo {year} {2007})}\BibitemShut {NoStop}%
\bibitem [{\citenamefont {Mitarai}\ \emph {et~al.}(2019)\citenamefont
  {Mitarai}, \citenamefont {Kitagawa},\ and\ \citenamefont
  {Fujii}}]{mitarai2019}%
  \BibitemOpen
  \bibfield  {author} {\bibinfo {author} {\bibfnamefont {K.}~\bibnamefont
  {Mitarai}}, \bibinfo {author} {\bibfnamefont {M.}~\bibnamefont {Kitagawa}}, \
  and\ \bibinfo {author} {\bibfnamefont {K.}~\bibnamefont {Fujii}},\
  }\href@noop {} {\bibfield  {journal} {\bibinfo  {journal} {Physical Review
  A}\ }\textbf {\bibinfo {volume} {99}},\ \bibinfo {pages} {012301} (\bibinfo
  {year} {2019})}\BibitemShut {NoStop}%
\bibitem [{\citenamefont {H{\"a}ner}\ \emph {et~al.}(2018)\citenamefont
  {H{\"a}ner}, \citenamefont {Roetteler},\ and\ \citenamefont
  {Svore}}]{haner2018}%
  \BibitemOpen
  \bibfield  {author} {\bibinfo {author} {\bibfnamefont {T.}~\bibnamefont
  {H{\"a}ner}}, \bibinfo {author} {\bibfnamefont {M.}~\bibnamefont
  {Roetteler}}, \ and\ \bibinfo {author} {\bibfnamefont {K.~M.}\ \bibnamefont
  {Svore}},\ }\href@noop {} {\bibfield  {journal} {\bibinfo  {journal} {arXiv
  preprint arXiv:1805.12445}\ } (\bibinfo {year} {2018})}\BibitemShut {NoStop}%
\bibitem [{\citenamefont {Woerner}\ and\ \citenamefont
  {Egger}(2019)}]{woerner2019}%
  \BibitemOpen
  \bibfield  {author} {\bibinfo {author} {\bibfnamefont {S.}~\bibnamefont
  {Woerner}}\ and\ \bibinfo {author} {\bibfnamefont {D.~J.}\ \bibnamefont
  {Egger}},\ }\href@noop {} {\bibfield  {journal} {\bibinfo  {journal} {npj
  Quantum Information}\ }\textbf {\bibinfo {volume} {5}},\ \bibinfo {pages} {1}
  (\bibinfo {year} {2019})}\BibitemShut {NoStop}%
\bibitem [{\citenamefont {Stamatopoulos}\ \emph {et~al.}(2019)\citenamefont
  {Stamatopoulos}, \citenamefont {Egger}, \citenamefont {Sun}, \citenamefont
  {Zoufal}, \citenamefont {Iten}, \citenamefont {Shen},\ and\ \citenamefont
  {Woerner}}]{stamatopoulos2019}%
  \BibitemOpen
  \bibfield  {author} {\bibinfo {author} {\bibfnamefont {N.}~\bibnamefont
  {Stamatopoulos}}, \bibinfo {author} {\bibfnamefont {D.~J.}\ \bibnamefont
  {Egger}}, \bibinfo {author} {\bibfnamefont {Y.}~\bibnamefont {Sun}}, \bibinfo
  {author} {\bibfnamefont {C.}~\bibnamefont {Zoufal}}, \bibinfo {author}
  {\bibfnamefont {R.}~\bibnamefont {Iten}}, \bibinfo {author} {\bibfnamefont
  {N.}~\bibnamefont {Shen}}, \ and\ \bibinfo {author} {\bibfnamefont
  {S.}~\bibnamefont {Woerner}},\ }\href@noop {} {\bibfield  {journal} {\bibinfo
   {journal} {arXiv preprint arXiv:1905.02666}\ } (\bibinfo {year}
  {2019})}\BibitemShut {NoStop}%
\bibitem [{\citenamefont {Yonehara}\ and\ \citenamefont
  {Takatsuka}(2010)}]{yonehara2010}%
  \BibitemOpen
  \bibfield  {author} {\bibinfo {author} {\bibfnamefont {T.}~\bibnamefont
  {Yonehara}}\ and\ \bibinfo {author} {\bibfnamefont {K.}~\bibnamefont
  {Takatsuka}},\ }\href@noop {} {\bibfield  {journal} {\bibinfo  {journal} {The
  Journal of Chemical Physics}\ }\textbf {\bibinfo {volume} {132}},\ \bibinfo
  {pages} {244102} (\bibinfo {year} {2010})}\BibitemShut {NoStop}%
\bibitem [{\citenamefont {Peruzzo}\ \emph {et~al.}(2014)\citenamefont
  {Peruzzo}, \citenamefont {McClean}, \citenamefont {Shadbolt}, \citenamefont
  {Yung}, \citenamefont {Zhou}, \citenamefont {Love}, \citenamefont
  {Aspuru-Guzik},\ and\ \citenamefont {O'Brien}}]{peruzzo2014}%
  \BibitemOpen
  \bibfield  {author} {\bibinfo {author} {\bibfnamefont {A.}~\bibnamefont
  {Peruzzo}}, \bibinfo {author} {\bibfnamefont {J.}~\bibnamefont {McClean}},
  \bibinfo {author} {\bibfnamefont {P.}~\bibnamefont {Shadbolt}}, \bibinfo
  {author} {\bibfnamefont {M.-H.}\ \bibnamefont {Yung}}, \bibinfo {author}
  {\bibfnamefont {X.-Q.}\ \bibnamefont {Zhou}}, \bibinfo {author}
  {\bibfnamefont {P.~J.}\ \bibnamefont {Love}}, \bibinfo {author}
  {\bibfnamefont {A.}~\bibnamefont {Aspuru-Guzik}}, \ and\ \bibinfo {author}
  {\bibfnamefont {J.~L.}\ \bibnamefont {O'Brien}},\ }\href@noop {} {\bibfield
  {journal} {\bibinfo  {journal} {Nature Communications}\ }\textbf {\bibinfo
  {volume} {5}} (\bibinfo {year} {2014})}\BibitemShut {NoStop}%
\bibitem [{\citenamefont {Yung}\ \emph {et~al.}(2014)\citenamefont {Yung},
  \citenamefont {Casanova}, \citenamefont {Mezzacapo}, \citenamefont {Mcclean},
  \citenamefont {Lamata}, \citenamefont {Aspuru-Guzik},\ and\ \citenamefont
  {Solano}}]{yung2014}%
  \BibitemOpen
  \bibfield  {author} {\bibinfo {author} {\bibfnamefont {M.-H.}\ \bibnamefont
  {Yung}}, \bibinfo {author} {\bibfnamefont {J.}~\bibnamefont {Casanova}},
  \bibinfo {author} {\bibfnamefont {A.}~\bibnamefont {Mezzacapo}}, \bibinfo
  {author} {\bibfnamefont {J.}~\bibnamefont {Mcclean}}, \bibinfo {author}
  {\bibfnamefont {L.}~\bibnamefont {Lamata}}, \bibinfo {author} {\bibfnamefont
  {A.}~\bibnamefont {Aspuru-Guzik}}, \ and\ \bibinfo {author} {\bibfnamefont
  {E.}~\bibnamefont {Solano}},\ }\href@noop {} {\bibfield  {journal} {\bibinfo
  {journal} {Scientific Reports}\ }\textbf {\bibinfo {volume} {4}},\ \bibinfo
  {pages} {3589} (\bibinfo {year} {2014})}\BibitemShut {NoStop}%
\bibitem [{\citenamefont {McClean}\ \emph {et~al.}(2016)\citenamefont
  {McClean}, \citenamefont {Romero}, \citenamefont {Babbush},\ and\
  \citenamefont {Aspuru-Guzik}}]{mcclean2016}%
  \BibitemOpen
  \bibfield  {author} {\bibinfo {author} {\bibfnamefont {J.~R.}\ \bibnamefont
  {McClean}}, \bibinfo {author} {\bibfnamefont {J.}~\bibnamefont {Romero}},
  \bibinfo {author} {\bibfnamefont {R.}~\bibnamefont {Babbush}}, \ and\
  \bibinfo {author} {\bibfnamefont {A.}~\bibnamefont {Aspuru-Guzik}},\
  }\href@noop {} {\bibfield  {journal} {\bibinfo  {journal} {New Journal of
  Physics}\ }\textbf {\bibinfo {volume} {18}},\ \bibinfo {pages} {023023}
  (\bibinfo {year} {2016})}\BibitemShut {NoStop}%
\bibitem [{\citenamefont {Wang}\ \emph {et~al.}(2019)\citenamefont {Wang},
  \citenamefont {Higgott},\ and\ \citenamefont {Brierley}}]{wang2019}%
  \BibitemOpen
  \bibfield  {author} {\bibinfo {author} {\bibfnamefont {D.}~\bibnamefont
  {Wang}}, \bibinfo {author} {\bibfnamefont {O.}~\bibnamefont {Higgott}}, \
  and\ \bibinfo {author} {\bibfnamefont {S.}~\bibnamefont {Brierley}},\
  }\href@noop {} {\bibfield  {journal} {\bibinfo  {journal} {Physical Review
  Letters}\ }\textbf {\bibinfo {volume} {122}},\ \bibinfo {pages} {140504}
  (\bibinfo {year} {2019})}\BibitemShut {NoStop}%
\bibitem [{\citenamefont {Grover}\ and\ \citenamefont
  {Rudolph}(2002)}]{grover2002creating}%
  \BibitemOpen
  \bibfield  {author} {\bibinfo {author} {\bibfnamefont {L.}~\bibnamefont
  {Grover}}\ and\ \bibinfo {author} {\bibfnamefont {T.}~\bibnamefont
  {Rudolph}},\ }\href@noop {} {\bibfield  {journal} {\bibinfo  {journal} {arXiv
  preprint quant-ph/0208112}\ } (\bibinfo {year} {2002})}\BibitemShut {NoStop}%
\bibitem [{\citenamefont {Kitaev}\ and\ \citenamefont
  {Webb}(2008)}]{kitaev2008wavefunction}%
  \BibitemOpen
  \bibfield  {author} {\bibinfo {author} {\bibfnamefont {A.}~\bibnamefont
  {Kitaev}}\ and\ \bibinfo {author} {\bibfnamefont {W.~A.}\ \bibnamefont
  {Webb}},\ }\href@noop {} {\bibfield  {journal} {\bibinfo  {journal} {arXiv
  preprint arXiv:0801.0342}\ } (\bibinfo {year} {2008})}\BibitemShut {NoStop}%
\bibitem [{\citenamefont {Capano}\ \emph {et~al.}(2014)\citenamefont {Capano},
  \citenamefont {Chergui}, \citenamefont {Rothlisberger}, \citenamefont
  {Tavernelli},\ and\ \citenamefont {Penfold}}]{capano2014}%
  \BibitemOpen
  \bibfield  {author} {\bibinfo {author} {\bibfnamefont {G.}~\bibnamefont
  {Capano}}, \bibinfo {author} {\bibfnamefont {M.}~\bibnamefont {Chergui}},
  \bibinfo {author} {\bibfnamefont {U.}~\bibnamefont {Rothlisberger}}, \bibinfo
  {author} {\bibfnamefont {I.}~\bibnamefont {Tavernelli}}, \ and\ \bibinfo
  {author} {\bibfnamefont {T.~J.}\ \bibnamefont {Penfold}},\ }\href@noop {}
  {\bibfield  {journal} {\bibinfo  {journal} {The Journal of Physical Chemistry
  A}\ }\textbf {\bibinfo {volume} {118}},\ \bibinfo {pages} {9861} (\bibinfo
  {year} {2014})}\BibitemShut {NoStop}%
\bibitem [{\citenamefont {Zhugayevych}\ and\ \citenamefont
  {Tretiak}(2015)}]{Tretiak_rev2015}%
  \BibitemOpen
  \bibfield  {author} {\bibinfo {author} {\bibfnamefont {A.}~\bibnamefont
  {Zhugayevych}}\ and\ \bibinfo {author} {\bibfnamefont {S.}~\bibnamefont
  {Tretiak}},\ }\href@noop {} {\bibfield  {journal} {\bibinfo  {journal}
  {Annual Review of Physical Chemistry}\ }\textbf {\bibinfo {volume} {66}},\
  \bibinfo {pages} {305} (\bibinfo {year} {2015})}\BibitemShut {NoStop}%
\end{thebibliography}

\begin{thebibliography}{11}%
\makeatletter
\providecommand \@ifxundefined [1]{%
 \@ifx{#1\undefined}
}%
\providecommand \@ifnum [1]{%
 \ifnum #1\expandafter \@firstoftwo
 \else \expandafter \@secondoftwo
 \fi
}%
\providecommand \@ifx [1]{%
 \ifx #1\expandafter \@firstoftwo
 \else \expandafter \@secondoftwo
 \fi
}%
\providecommand \natexlab [1]{#1}%
\providecommand \enquote  [1]{``#1''}%
\providecommand \bibnamefont  [1]{#1}%
\providecommand \bibfnamefont [1]{#1}%
\providecommand \citenamefont [1]{#1}%
\providecommand \href@noop [0]{\@secondoftwo}%
\providecommand \href [0]{\begingroup \@sanitize@url \@href}%
\providecommand \@href[1]{\@@startlink{#1}\@@href}%
\providecommand \@@href[1]{\endgroup#1\@@endlink}%
\providecommand \@sanitize@url [0]{\catcode `\\12\catcode `\$12\catcode
  `\&12\catcode `\#12\catcode `\^12\catcode `\_12\catcode `\%12\relax}%
\providecommand \@@startlink[1]{}%
\providecommand \@@endlink[0]{}%
\providecommand \url  [0]{\begingroup\@sanitize@url \@url }%
\providecommand \@url [1]{\endgroup\@href {#1}{\urlprefix }}%
\providecommand \urlprefix  [0]{URL }%
\providecommand \Eprint [0]{\href }%
\providecommand \doibase [0]{http://dx.doi.org/}%
\providecommand \selectlanguage [0]{\@gobble}%
\providecommand \bibinfo  [0]{\@secondoftwo}%
\providecommand \bibfield  [0]{\@secondoftwo}%
\providecommand \translation [1]{[#1]}%
\providecommand \BibitemOpen [0]{}%
\providecommand \bibitemStop [0]{}%
\providecommand \bibitemNoStop [0]{.\EOS\space}%
\providecommand \EOS [0]{\spacefactor3000\relax}%
\providecommand \BibitemShut  [1]{\csname bibitem#1\endcsname}%
\let\auto@bib@innerbib\@empty
\bibitem [{\citenamefont {Berry}\ \emph {et~al.}(2007)\citenamefont {Berry},
  \citenamefont {Ahokas}, \citenamefont {Cleve},\ and\ \citenamefont
  {Sanders}}]{berry2007}%
  \BibitemOpen
  \bibfield  {author} {\bibinfo {author} {\bibfnamefont {D.~W.}\ \bibnamefont
  {Berry}}, \bibinfo {author} {\bibfnamefont {G.}~\bibnamefont {Ahokas}},
  \bibinfo {author} {\bibfnamefont {R.}~\bibnamefont {Cleve}}, \ and\ \bibinfo
  {author} {\bibfnamefont {B.~C.}\ \bibnamefont {Sanders}},\ }\href@noop {}
  {\bibfield  {journal} {\bibinfo  {journal} {Communications in Mathematical
  Physics}\ }\textbf {\bibinfo {volume} {270}},\ \bibinfo {pages} {359}
  (\bibinfo {year} {2007})}\BibitemShut {NoStop}%
\bibitem [{\citenamefont {Benenti}\ and\ \citenamefont
  {Strini}(2008)}]{benenti2008}%
  \BibitemOpen
  \bibfield  {author} {\bibinfo {author} {\bibfnamefont {G.}~\bibnamefont
  {Benenti}}\ and\ \bibinfo {author} {\bibfnamefont {G.}~\bibnamefont
  {Strini}},\ }\href@noop {} {\bibfield  {journal} {\bibinfo  {journal}
  {American Journal of Physics}\ }\textbf {\bibinfo {volume} {76}},\ \bibinfo
  {pages} {657} (\bibinfo {year} {2008})}\BibitemShut {NoStop}%
\bibitem [{\citenamefont {Stamatopoulos}\ \emph {et~al.}(2019)\citenamefont
  {Stamatopoulos}, \citenamefont {Egger}, \citenamefont {Sun}, \citenamefont
  {Zoufal}, \citenamefont {Iten}, \citenamefont {Shen},\ and\ \citenamefont
  {Woerner}}]{stamatopoulos2019}%
  \BibitemOpen
  \bibfield  {author} {\bibinfo {author} {\bibfnamefont {N.}~\bibnamefont
  {Stamatopoulos}}, \bibinfo {author} {\bibfnamefont {D.~J.}\ \bibnamefont
  {Egger}}, \bibinfo {author} {\bibfnamefont {Y.}~\bibnamefont {Sun}}, \bibinfo
  {author} {\bibfnamefont {C.}~\bibnamefont {Zoufal}}, \bibinfo {author}
  {\bibfnamefont {R.}~\bibnamefont {Iten}}, \bibinfo {author} {\bibfnamefont
  {N.}~\bibnamefont {Shen}}, \ and\ \bibinfo {author} {\bibfnamefont
  {S.}~\bibnamefont {Woerner}},\ }\href@noop {} {\bibfield  {journal} {\bibinfo
   {journal} {arXiv preprint arXiv:1905.02666}\ } (\bibinfo {year}
  {2019})}\BibitemShut {NoStop}%
\bibitem [{\citenamefont {Yonehara}\ and\ \citenamefont
  {Takatsuka}(2010)}]{yonehara2010}%
  \BibitemOpen
  \bibfield  {author} {\bibinfo {author} {\bibfnamefont {T.}~\bibnamefont
  {Yonehara}}\ and\ \bibinfo {author} {\bibfnamefont {K.}~\bibnamefont
  {Takatsuka}},\ }\href@noop {} {\bibfield  {journal} {\bibinfo  {journal} {The
  Journal of Chemical Physics}\ }\textbf {\bibinfo {volume} {132}},\ \bibinfo
  {pages} {244102} (\bibinfo {year} {2010})}\BibitemShut {NoStop}%
\bibitem [{\citenamefont {Peruzzo}\ \emph {et~al.}(2014)\citenamefont
  {Peruzzo}, \citenamefont {McClean}, \citenamefont {Shadbolt}, \citenamefont
  {Yung}, \citenamefont {Zhou}, \citenamefont {Love}, \citenamefont
  {Aspuru-Guzik},\ and\ \citenamefont {O'Brien}}]{peruzzo2014}%
  \BibitemOpen
  \bibfield  {author} {\bibinfo {author} {\bibfnamefont {A.}~\bibnamefont
  {Peruzzo}}, \bibinfo {author} {\bibfnamefont {J.}~\bibnamefont {McClean}},
  \bibinfo {author} {\bibfnamefont {P.}~\bibnamefont {Shadbolt}}, \bibinfo
  {author} {\bibfnamefont {M.-H.}\ \bibnamefont {Yung}}, \bibinfo {author}
  {\bibfnamefont {X.-Q.}\ \bibnamefont {Zhou}}, \bibinfo {author}
  {\bibfnamefont {P.~J.}\ \bibnamefont {Love}}, \bibinfo {author}
  {\bibfnamefont {A.}~\bibnamefont {Aspuru-Guzik}}, \ and\ \bibinfo {author}
  {\bibfnamefont {J.~L.}\ \bibnamefont {O'Brien}},\ }\href@noop {} {\bibfield
  {journal} {\bibinfo  {journal} {Nature Communications}\ }\textbf {\bibinfo
  {volume} {5}} (\bibinfo {year} {2014})}\BibitemShut {NoStop}%
\bibitem [{\citenamefont {Yung}\ \emph {et~al.}(2014)\citenamefont {Yung},
  \citenamefont {Casanova}, \citenamefont {Mezzacapo}, \citenamefont {Mcclean},
  \citenamefont {Lamata}, \citenamefont {Aspuru-Guzik},\ and\ \citenamefont
  {Solano}}]{yung2014}%
  \BibitemOpen
  \bibfield  {author} {\bibinfo {author} {\bibfnamefont {M.-H.}\ \bibnamefont
  {Yung}}, \bibinfo {author} {\bibfnamefont {J.}~\bibnamefont {Casanova}},
  \bibinfo {author} {\bibfnamefont {A.}~\bibnamefont {Mezzacapo}}, \bibinfo
  {author} {\bibfnamefont {J.}~\bibnamefont {Mcclean}}, \bibinfo {author}
  {\bibfnamefont {L.}~\bibnamefont {Lamata}}, \bibinfo {author} {\bibfnamefont
  {A.}~\bibnamefont {Aspuru-Guzik}}, \ and\ \bibinfo {author} {\bibfnamefont
  {E.}~\bibnamefont {Solano}},\ }\href@noop {} {\bibfield  {journal} {\bibinfo
  {journal} {Scientific Reports}\ }\textbf {\bibinfo {volume} {4}},\ \bibinfo
  {pages} {3589} (\bibinfo {year} {2014})}\BibitemShut {NoStop}%
\bibitem [{\citenamefont {McClean}\ \emph {et~al.}(2016)\citenamefont
  {McClean}, \citenamefont {Romero}, \citenamefont {Babbush},\ and\
  \citenamefont {Aspuru-Guzik}}]{mcclean2016}%
  \BibitemOpen
  \bibfield  {author} {\bibinfo {author} {\bibfnamefont {J.~R.}\ \bibnamefont
  {McClean}}, \bibinfo {author} {\bibfnamefont {J.}~\bibnamefont {Romero}},
  \bibinfo {author} {\bibfnamefont {R.}~\bibnamefont {Babbush}}, \ and\
  \bibinfo {author} {\bibfnamefont {A.}~\bibnamefont {Aspuru-Guzik}},\
  }\href@noop {} {\bibfield  {journal} {\bibinfo  {journal} {New Journal of
  Physics}\ }\textbf {\bibinfo {volume} {18}},\ \bibinfo {pages} {023023}
  (\bibinfo {year} {2016})}\BibitemShut {NoStop}%
\bibitem [{\citenamefont {Wang}\ \emph {et~al.}(2019)\citenamefont {Wang},
  \citenamefont {Higgott},\ and\ \citenamefont {Brierley}}]{wang2019}%
  \BibitemOpen
  \bibfield  {author} {\bibinfo {author} {\bibfnamefont {D.}~\bibnamefont
  {Wang}}, \bibinfo {author} {\bibfnamefont {O.}~\bibnamefont {Higgott}}, \
  and\ \bibinfo {author} {\bibfnamefont {S.}~\bibnamefont {Brierley}},\
  }\href@noop {} {\bibfield  {journal} {\bibinfo  {journal} {Physical Review
  Letters}\ }\textbf {\bibinfo {volume} {122}},\ \bibinfo {pages} {140504}
  (\bibinfo {year} {2019})}\BibitemShut {NoStop}%
\bibitem [{\citenamefont {Spall}(2000)}]{spall2000}%
  \BibitemOpen
  \bibfield  {author} {\bibinfo {author} {\bibfnamefont {J.~C.}\ \bibnamefont
  {Spall}},\ }\href {\doibase 10.1109/TAC.2000.880982} {\bibfield  {journal}
  {\bibinfo  {journal} {IEEE Transactions on Automatic Control}\ }\textbf
  {\bibinfo {volume} {45}},\ \bibinfo {pages} {1839} (\bibinfo {year}
  {2000})}\BibitemShut {NoStop}%
\bibitem [{\citenamefont {Marcus}(1956)}]{marcus1956}%
  \BibitemOpen
  \bibfield  {author} {\bibinfo {author} {\bibfnamefont {R.~A.}\ \bibnamefont
  {Marcus}},\ }\href@noop {} {\bibfield  {journal} {\bibinfo  {journal} {The
  Journal of chemical physics}\ }\textbf {\bibinfo {volume} {24}},\ \bibinfo
  {pages} {966} (\bibinfo {year} {1956})}\BibitemShut {NoStop}%
\bibitem [{\citenamefont {Nitzan}(2006)}]{nitzan2006}%
  \BibitemOpen
  \bibfield  {author} {\bibinfo {author} {\bibfnamefont {A.}~\bibnamefont
  {Nitzan}},\ }\href@noop {} {\emph {\bibinfo {title} {Chemical dynamics in
  condensed phases: relaxation, transfer and reactions in condensed molecular
  systems}}}\ (\bibinfo  {publisher} {Oxford university press},\ \bibinfo
  {year} {2006})\BibitemShut {NoStop}%
\end{thebibliography}
\end{document}